\documentstyle{mn}
\input{epsf}
\newcommand{\be}{\begin{eqnarray}}
\newcommand{\ee}{\end{eqnarray}}

\title[Characteristics of Ly-$\alpha$ forest ]
{Statistical characteristics  of the observed Ly-$\alpha$ forest\\
and the shape of the initial power spectrum}
\author[Demia\'nski,  Doroshkevich \& Turchaninov]
       {M. Demia\'nski$^{1,2}$,  A.G. Doroshkevich$^{3}$,
V.I.Turchaninov$^4$\\
        $1$Institute of Theoretical Physics,
                       University of Warsaw,
                       00-681 Warsaw, Poland\\
        $2$Department of Astronomy, Williams College,
           Williamstown, MA 01267, USA\\
        $3$Astro Space Center of Lebedev Physical
           Institute of  Russian Academy of Sciences,
                        117997 Moscow,  Russia\\
        $4$ Keldysh Institute of  Applied Mathematics
             Russian Academy of Sciences,
                        125047 Moscow,  Russia\\
}

\date{Accepted ...,
      Received ...,
        in original form ... .}

\begin{document}
\maketitle

\begin{abstract}
We analyze basic properties of about 6000 Lyman-$\alpha$
absorbers observed in high resolution spectra of 19 quasars. We
compare their observed characteristics with predictions of our
model of formation and evolution of absorbers and dark matter
(DM) pancakes and voids based on the Zel'dovich theory of
gravitational instability. This model asserts that absorbers are
formed in the course of both linear and non linear adiabatic and
shock compression of DM and gaseous matter. Our model is
consistent with simulations of structure formation, describes
reasonably well the Large Scale Structure (LSS) observed in the
distribution of galaxies at small redshifts and emphasizes the
generic similarity of the process of formation of LSS and
absorbers. Using this model we are able to link the column
density and overdensity of DM and gaseous component with the
observed column density of neutral hydrogen, redshifts and
Doppler parameters of absorbers. We show that the colder
absorbers are associated with rapidly expanded underdense
regions of galactic scale.

We extend the method of measuring the power spectrum
of initial perturbations proposed in Demia\'nski \& Doroshkevich
(2003b). The observed separations between absorbers and their DM
column density are linked with the correlation function of the
initial velocity field. Applying this method to our sample of
absorbers we recover the CDM like power spectrum at scales
$10h^{-1}\geq D\geq 0.15h^{-1}$ Mpc with a precision of $\sim$
15\%. However at scales $\sim 3 - 150 h^{-1}$kpc the measured
and CDM--like spectra are different. This result suggests a
possible complex inflation with generation of excess power at
small scales. Both confirmation of the CDM--like shape of the
initial power spectrum or detection of its distortions at small
scales are equally important for the widely discussed problems
of physics of the early Universe, galaxy formation, and
reheating of the Universe.
\end{abstract}


\section{Introduction}
One of the most perspective methods to study the processes
responsible for the formation and evolution of the structure in
the Universe is the analysis of properties of absorbers observed
in spectra of the farthest quasars. The great potential of such
investigations was discussed already by Oort (1981, 1984) just
after Sargent et al. (1980) established the intergalactic nature
of the Lyman-$\alpha$ forest. Indeed, the absorption lines trace
the small scale distribution of hydrogen along the line of sight
at redshifts $z\geq$ 2 when matter is not yet strongly clustered
and its observed characteristics can be more easily interpreted.
The available Keck and VLT high resolution observations of the
Lyman-$\alpha$ forest provide a reasonable database and allow
one to apply statistical methods for their analysis.

The composition and spatial distribution of the observed
absorbers is complicated. Thus, at large redshifts the
population of rich metal systems including Ly-damped and
Ly-limit systems are rare and majority of observed absorbers is
associated with isolated low mass HI clouds. At low redshifts a
significant number of stronger Ly-$\alpha$ lines and metal
systems is associated with galaxies (Bergeron et al. 1992;
Lanzetta et al. 1995; Tytler 1995; Le Brune et al. 1996).
However as was recently shown by Penton, Stock and Shull (2000;
2002) and McLin et al. (2002), even at small redshifts some
absorbers are associated with galaxy filaments while others are
found within galaxy voids. These results suggest that the
population of weaker absorbers dominating at higher redshifts
can be associated with weaker structure elements formed by the
non luminous baryonic and DM components.
They also suggest that the Ly--$\alpha$ forest can be
considered as a low mass component of the generic Large Scale
Structure (LSS) which is
seen in simulated and observed spatial matter distribution.
This means that absorbers at high $z$ trace the DM structure
which is qualitatively similar to the rescaled one observed at
small redshifts.

In this paper we investigate a sample of $\sim 6000$ absorbers
observed in 19 high resolution spectra of QSOs and compare their
properties with the improved model of absorbers proposed in
Demia\'nski \&  Doroshkevich (2003 a,b, hereafter Paper I \& Paper
II). We assume that absorbers are dominated by long--lived
gravitationally bound and partially relaxed clouds composed of both
DM and baryonic components. The fact that we can observe galaxies
and quasars at $z\geq 3$ demonstrates the existence of strong
density perturbations already then. Here we show that at these
redshifts there are also strong negative density perturbations of a
galactic scale which can be identified with rapidly expanded
underdense regions. Such regions are naturally associated with the
colder absorbers.

Our model explains the self similar character of evolution of the
observed Doppler parameter, HI column density and absorber
separation. Such evolution implies that the mean values of these
characteristics slowly vary with redshift while their probability
distribution functions (PDFs) remain unchanged. This model links the
observed and other physical characteristics of absorbers -- such as
their DM column density, size, and fraction of matter associated with
absorbers -- and allows us also to identify several subpopulations
of absorbers with different evolutionary histories.

We treat the evolution of structure as a random process of
formation and merging of Zel'dovich pancakes, their transverse
expansion and/or compression and successive transformation into
filaments and high density clouds. Later on the hierarchical
merging of pancakes, filaments and clouds forms rich galaxy
walls observed at small redshifts. Impact of these factors is
clearly seen in high resolution numerical simulations of
evolution of the LSS (for review see,
e.g. Frenk 2002).

Theoretical expectations of our model are based on the Zel'dovich
approximate theory of gravitational instability (Zel'dovich 1970;
Shandarin \& Zel'dovich 1988). As is well known, it
correctly describes only the linear and weakly nonlinear stages
of the structure formation and cannot describe later stages of
evolution of structure elements. In spite of this, the statistical
approach proposed in (Demia\'nski \& Doroshkevich 1999, 2004a;
hereafter DD99 \& DD04) nicely describes the main properties of
observed and simulated LSS (Demia\'nski et al. 2000; Doroshkevich,
Tucker, Allam \& Way 2004) without any smoothing or filtering
procedure.

Presently various observations are used to determine the power
spectrum of the initial density perturbations. Its amplitude and 
its shape on scales $\geq 10h^{-1}$Mpc are approximately established 
by investigations of the microwave relic radiation (Spergel et al.
2003, 2006) and the structure  of the Universe at $z\ll$~1 detected 
in large redshift surveys (Percival et al. 2001; Tegmark, Hamilton 
\& Xu 2003; Verde et al. 2002, 2003) and weak lensing data (see, 
e.g., Hoekstra, Yee \& Gladders 2002).

The shape of the initial power spectrum on scales $10h^{-1}$ Mpc --
$1h^{-1}$Mpc can be tested at high redshifts where it is not yet
strongly distorted by nonlinear evolution (Croft et al. 1998, 2002;
Nusser  \& Haehnelt 2000; Gnedin \& Hamilton 2002; Viel et al. 2004
a,b; Kim et al. 2004; McDonald et al. 2004; Seljak  et al. 2004;
Zaroubi et al. 2005).
The method used in these papers is surprisingly universal and is
successfully applied to spectra observed with both high and moderate
resolution. It utilizes the measured transmitted flux only and does
not require preliminary determination of column density, Doppler
parameters and even discrimination of hydrogen and metal line
systems (McDonald et al. 2004, Seljak  et al. 2004). In spite of
this, it successfully restores the CDM--like power spectrum down to
scales $\sim 1h^{-1}$ Mpc. Recent results on reconstruction of the
initial power spectrum are summarized and discussed in many papers
(see, e.g., Tegmark and Zaldarriaga 2003; Wang et al. 2003;
Zaldarriaga, Scoccimorro \& Hui 2003; Peiris et al. 2003; Spergel 
et al. 2003,2006; Tegmark et al. 2004; McDonald et al. 2004, Seljak 
et al. 2004).

A straightforward method of reconstruction of the initial power
spectrum from the observed characteristics of absorbers was proposed
and tested in Paper II. This method can be used to recover the
initial power spectrum down to unprecedentedly small scale. In
contrast with previous investigations (Croft et al. 1998, 2002;
Nusser \& Haehnelt 2000; Viel et al. 2004b; McDonald et al. 2004) we
analyze the separation between adjacent absorbers and their column
density rather than the flux or smoothed density field. This means
that our results are not restricted by the standard factors such as
the Nyquist limit, the impact of nonlinear processes, the unknown
matter distribution between absorbers or their peculiar velocities.
This approach successfully complements investigations of the power
spectrum mentioned above.

Here we improve the analysis discussed in Paper II by using a
richer observed sample and a more refined model of
absorbers. We use two independent methods of determination of the
initial power spectrum. The first one is based on measurements of
the separation between adjacent absorbers, while the second one,
proposed in Paper II, uses measurements of the column density of
absorbers. Both approaches allow one to determine the spectrum down
to the scale of $\sim 5 - 10 h^{-1}$ kpc. At scales $(0.15 -
10)h^{-1}$ Mpc our results coincide with those expected for the
CDM--like power spectrum and Gaussian perturbations with the
precision of $\sim$ 15\%. However, we have found some evidence that
at scales $\leq 0.15 h^{-1}$Mpc the initial power spectrum differs
from the CDM--like one suggesting a complex inflation with
generation of excess power at small scales. Such excess power
accelerates the process of galaxy formation at high redshifts and
can shift the epoch of reheating of the Universe to higher redshifts.

At present we have only limited information about the properties of
the background gas and the UV radiation (Haardt \& Madau 1996; Scott
et al. 2000, 2002; Schaye et al. 2000; McDonald \& Miralda--Escude
2001; McDonald et al. 2000, 2001; Theuns et al. 2002 a, b; Levshakov
at al. 2003; Boksenberg, Sargent \& Rauch 2003; Demia\'nski \&
Doroshkevich 2004b) and therefor some numerical factors in our model
remain undetermined. This means that our approach should be tested
on representative numerical simulations that more accurately follow
the process of formation and disruption of pancakes and filaments
and provide a unified picture of the process of absorbers formation
and evolution (see, e.g., Weinberg et al. 1998; Zhang et al. 1998;
Dav\'e et al. 1999; Theuns et al. 1999, 2000). But so far such
simulations
are performed mainly in small boxes what restricts their
representativity, introduces artificial cutoffs in the power
spectrum and complicates the quantitative description of structure
evolution (see more detailed discussion in Gnedin \& Hamilton 2002;
Tegmark \& Zaldarriaga 2002; Zaldarriaga Scoccimoro \& Hui 2002;
Seljak, McDonald \& Makarov 2003; Manning 2003; Paper II). As was
shown by Meiksin, Bryan \& Machacek (2001), the available simulations
reproduce quite well the characteristics of the flux but cannot
restore other observed characteristics of the forest. This means
that first of all numerical simulations should be improved (see more
detailed discussion in Paper II).

Comparison of results obtained in Paper I and Paper II and in
this paper demonstrates that the quality and representativity of
the sample of observed absorbers are very important for the
reconstruction of processes of absorbers formation and
evolution. Thus, richer sample of the observed absorbers makes
it possible to select and investigate several representative
subsamples of absorbers with different evolutionary histories.
None the less, our analysis indicates a possible deficit of
weaker absorbers and pairs of close absorbers what in turn could
be related to insufficient sensitivity of the process of
absorbers' identification.  Thus, the number and parameters of
absorbers identified for the same quasar depend upon the used
identification procedure. Further progress can be achieved first
of all with richer samples covering the range of redshifts at
least up to $z\sim$~5 what would allow one to perform a
complex investigation of the early period of structure
evolution.

This paper is organized as follows. In Sec. 2 the observational
databases used in our analysis are presented and statistical
characteristics of the observed parameters of absorbers are
obtained. The theoretical models of the structure evolution are
discussed in Sec. 3. The results of statistical analysis of the
model dependent parameters and the derived initial power
spectrum are given in Sec. 4. Discussion and conclusions can be
found in Sec. 5.

\section{Observed characteristics of absorbers}

\subsection{Properties of the homogeneously distributed
matter}

In this paper we consider the spatially flat $\Lambda$CDM
model of the Universe with the Hubble parameter and mean
density given by:
\[
H^{2}(z) = H_0^2\Omega_m(1+z)^3[1+\Omega_\Lambda/\Omega_m
(1+z)^{-3}]\,,
\]
\be
\langle n_b(z)\rangle = 2.4\cdot 10^{-7}
(1+z)^3(\Omega_bh^2/0.02){\rm cm}^{-3}\,,
\label{basic}
\ee
\[
\langle\rho_m(z)\rangle = {3H_0^2\over 8\pi G}\Omega_m(1+z)^3,
\quad H_0=100h\,{\rm km/s/Mpc}\,.
\]
Here $\Omega_m=0.3\,\&\,\Omega_\Lambda=0.7$ are the
dimensionless matter density and the cosmological constant (dark
energy), $\Omega_b$ is the dimensionless mean density of
baryons, and $h=$~0.7 is the dimensionless Hubble constant.

Properties of the compressed gas can be suitably related to the
parameters of homogeneously distributed gas, which were discussed in
many papers (see, e.g., Ikeuchi\,\&\,Ostriker 1986; Haardt \& Madau
1996; Hui \,\&\,Gnedin 1997; Scott et al. 2000; McDonald et  al.
2001; Theuns et al. 2002a, b; Demia\'nski \& Doroshkevich 2004b).
In this paper we consider evolution of absorbers at observed
redshits $z\leq 4$ when weak variations of the gas entropy are
determined by the interaction of the gas with the UV background.
Thus, the expected intensity of the UV background radiation, can be
fitted as follows: \be J(z,\nu)=J_{21}(z)\left({\nu_H\over\nu}\right
)^{\alpha_\gamma} \cdot 10^{-21}{erg\over s\,cm^2\,sr\,Hz}\,,
\label{uvrad} \ee where $\nu_H=3.3\cdot 10^{15}$ Hz,
$\alpha_\gamma\approx 1.5$ and the dimensionless factor $J_{21}(z)$
describes redshift variations of the intensity. The mean
temperature, $\langle T_{bg}\rangle$, of homogeneous gas can be
taken as
\[
\langle T_{bg}\rangle\approx 3.5z_4^{6/7}\Theta_{bg}^{4/7}
\cdot 10^4K\,,
\]
\be
\langle b_{bg}\rangle=\sqrt{2k_B\langle T_{bg}\rangle \over
m_H}\approx 24 z_4^{3/7}\Theta_{bg}^{2/7}{\rm km/s}\,,
\label{bg}
\ee
\[
\Theta_{bg}={\Omega_bh^2\over 0.02}{3.5\over 2+\alpha_\gamma}
\left({0.15\over\Omega_mh^2}
\right)^{3/4},\quad z_4={1+z\over 4}\,,
\]
where $k_B\,\&\,m_H$ are the Boltzmann constant and the mass of the
hydrogen atom and $\alpha_\gamma$ is the power index of spectrum of
the ionizing background in (\ref{uvrad}). Analyzing the observed
characteristics of absorbers McDonald et al. (2001) estimate the
background temperature as $T_{bg}\approx (2\pm 0.2)\cdot 10^4 K$ at
$z\sim 2$, what is close to (\ref{bg}).

At this period the gas entropy can be characterized by the
function \be \langle F_{bg}\rangle={\langle
T_{bg}\rangle\over\langle n_b \rangle^{2/3}} = 60
z_4^{-8/7}\Theta_{bg}^{4/7}
\left({0.02\over\Omega_bh^2}\right)^{2/3}{\rm keV\cdot cm^2}\,.
\label{sbg} \ee As was shown in Demia\'nski \& Doroshkevich
(2004b), the function $F_{bg}\propto (n_b/\langle
n_b\rangle)^{0.1}$ only weakly depends upon variations of the
expansion rate. This means that the background temperature and
Doppler parameter vary as \be T_{bg}\propto (n_b/\langle
n_b\rangle)^{2/3},\quad b_{bg}\propto (n_b/\langle
n_b\rangle)^{1/3}\,, \label{rap} \ee and the mildly nonlinear
compression or expansion of mater occurs almost adiabatically.
Thus, within compressed or slowly expanded regions with
$n_b\ge\langle n_b \rangle$ we can expect that $T_{bg}\ge\langle
T_{bg}\rangle$\, while in rapidly expanded regions with
$n_b\ll\langle n_b \rangle$ we can expect that $T_{bg}\ll\langle
T_{bg}\rangle$\,.

Under the assumption of ionization equilibrium of the
gas,
\be
{\langle n_H\rangle\over\langle n_b\rangle}=
{\alpha_r\langle n_b\rangle\over\Gamma_\gamma},
\quad \alpha_r\approx 4.4\cdot 10^{-13}\left({10^4K
\over T}\right)^{3/4} {{\rm cm}^3\over{\rm  s}}\,,
\label{ieq}
\ee
where $\alpha_r(T)$ is the recombination coefficient
(Black, 1981) and $\Gamma_\gamma$ characterizes the
rate of ionization by the UV background, the fraction
of neutral hydrogen is
\be
x_{bg} = \langle n_H\rangle/\langle n_b\rangle =
x_0(1+z)^{33/14},
\label{xbg}
\ee
\[
x_0\approx {1.2\cdot 10^{-7}\over\Gamma_{12}(z)}{\Omega_bh^2
\over 0.02}\Theta_{bg}^{-3/7}, \quad \Gamma_\gamma(z)
=10^{-12}\Gamma_{12}(z)s^{-1}\,,
\]
\[
\Gamma_{12}(z) =12.6J_{21}(z)(3+\alpha_\gamma)^{-1}\,.
\]

The redshift variations of the rate of ionization,
$\Gamma_{12}(z)$, produced by the UV radiation of quasars were
discussed by Haardt \& Madau (1996) and later on tested and
corrected by Demia\'nski \& Doroshkevich (2004b) with large
observed sample of QSOs. For $\alpha_\gamma \approx 1.5$ the
expected ionization rate of hydrogen is fitted by the
expression: \be \Gamma_{12}\approx 7\exp[-(z-2.35)^2/2]\,.
\label{ioniz} \ee Variations of the power index $\alpha_\gamma$
with time and space generate random variations of $\Gamma_{12}$.
With these parameters of background we have for the
Gunn--Peterson optical depth \be \tau_{GP}(z)\approx {0.34
\over\Gamma_{12}\Theta_{bg}^{3/7}} \left({1+z\over
4}\right)^{27/7}\left({\Omega_bh^2\over 0.02}
\right)^2\sqrt{0.15\over\Omega_mh^2}\,, \label{gp1} \ee
\[
\tau_{GP}(2)\approx 0.015,\quad \tau_{GP}(5)\approx 8.2\,.
\]

However these estimates of $\Gamma_{12}$ should be corrected for
absorption and reemission of radiation within high density
clouds (Haardt \& Madau 1996). This effect decreases the
ionization rate of both hydrogen and helium and enhances a
possible spatial variations of intensity of the UV background.
According to the observational estimates of Scott et al. (2002)
$\Gamma_{12}\sim 1 - 4$ at $z\approx 2$ (see also Levshakov et
al. 2003; Boksenberg, Sargent\,\&\,Rauch 2003). On the other
hand, Eq. (\ref{ioniz}) takes into account only the contribution
of QSOs and, so, it underestimates the ionization rate at both
$z\leq 2$ and $z\geq 3$ where contribution of other probable
sources of radiation becomes important.

\begin{table}
\caption{QSO spectra used in our analysis} \label{tbl1}
\begin{tabular}{lll lr}
    &$z_{em}$&$z_{min}$&$z_{max}$&No \\
               &     &   &   &of HI lines\\
$1055+461^{1}~$ & 4.16&2.8&4.16& 998\\      
$0000-260^{2}~$ & 4.11&3.4&4.1& 431\\       
$0055-259^{3}~$ & 3.66&3.0&3.6& 534\\       
$1422+23^{1}~~~$& 3.6 &2.7&3.6& 811\\       
$0014+813^{4}~$ & 3.41&2.7&3.2& 262\\       
$0956+122^{4}~$ & 3.30&2.6&3.1& 256\\       
$0302-003^{4, 3}$&3.29&2.6&3.1& 356\\       
$0636+680^{1}~$ & 3.17&2.5&3.0& 531\\       
$0636+680^{4}~$ & 3.17&2.4&3.1& 313\\       
$1759+754^{5}~$ & 3.05&2.4&3.0& 307\\       
$1946+766^{6}~$ & 3.02&2.4&3.0& 461\\       
$1107+485^{1}~$ & 3.0 &2.1&3.0& 609\\       
$1347-246^{3}~$ & 2.63&2.1&2.6& 361\\       
$1122-441^{3}~$ & 2.42&1.9&2.4& 353\\       
$2217-282^{3}~$ & 2.41&1.9&2.3& 262\\       
$1626-643^{1}~$ & 2.32&1.5&2.3& 281\\       
$2233-606^{7}~$ & 2.24&1.5&2.2& 293\\       
$1101-264^{3}~$ & 2.15&1.6&2.1& 277\\       
$0515-441^{3}~$ & 1.72&1.5&1.7&~~76\\       
\vspace{0.15cm}
\end{tabular}                              

1. Rough \& Sargent, unpublished
2. Lu et al. (1996),
3. Kim et al. 2002
4. Hu et al., (1995),
5. Djorgovski et al. (2001)
6. Kirkman \& Tytler (1997),
7. Cristiani \& D'Odorico (2000),
\end{table}

\subsection{ The database.}

The present analysis is based on 19 high resolution spectra
listed in Table 1. These spectra contain 7\,770 absorbers.
For further discussion we selected the sample
of 7\,430 absorbers with $11.9\leq \lg N_{HI}\leq 15$ and 7\,411
distances between neighboring absorbers.  This sample will be
used for discussion of the correlation function of initial
velocity field. For detailed investigation we selected
a more homogeneous sample of 6\,270 absorbers and 6\,251
separations with $b\geq 5\,km/s$ and we restricted errors of
measurement by the conditions $\Delta \lg N_{HI}\leq 0.2$ and
$\Delta b\leq 0.3b$.  The chosen low limit of $b$ is close to
the spectral resolution in this sample. To test the sample
dependence of the correlation function of initial velocity
field we use for comparison the sample of 14 QSOs with 4\,036
absorbers investigated in Paper I and Paper II\,.

However, a list of absorbers depends also upon the method of
line identification and, for example, for QSO 0636+680 two
spectra listed in Table 1 include different number of absorbers.
This example shows that the methods of line identification
should be unified and improved. None the less, dispersions of
absorbers characteristics discussed below are defined mainly by
their broad distribution functions and by the completeness of
the samples. Because of this, in this paper we discuss the
scatter of only the more interesting quantitative
characteristics of absorbers.

As is seen from Fig. 1, the redshift distribution of
absorbers is non homogeneous and the majority of absorbers are
concentrated at 2~$\leq z\leq$~3.5\,. This means that some of
the discussed here characteristics of absorbers are derived
mainly from this range of redshifts. Absorbers at $z\geq~3.5$
were identified mainly in spectra of QSOs 0000-260 and 1055+461.
In this range and at $z\leq~2$ the statistics of lines is not
sufficient.

\subsection{Observed characteristics of absorbers}

\begin{figure}
\centering\epsfxsize=7.cm
\epsfbox{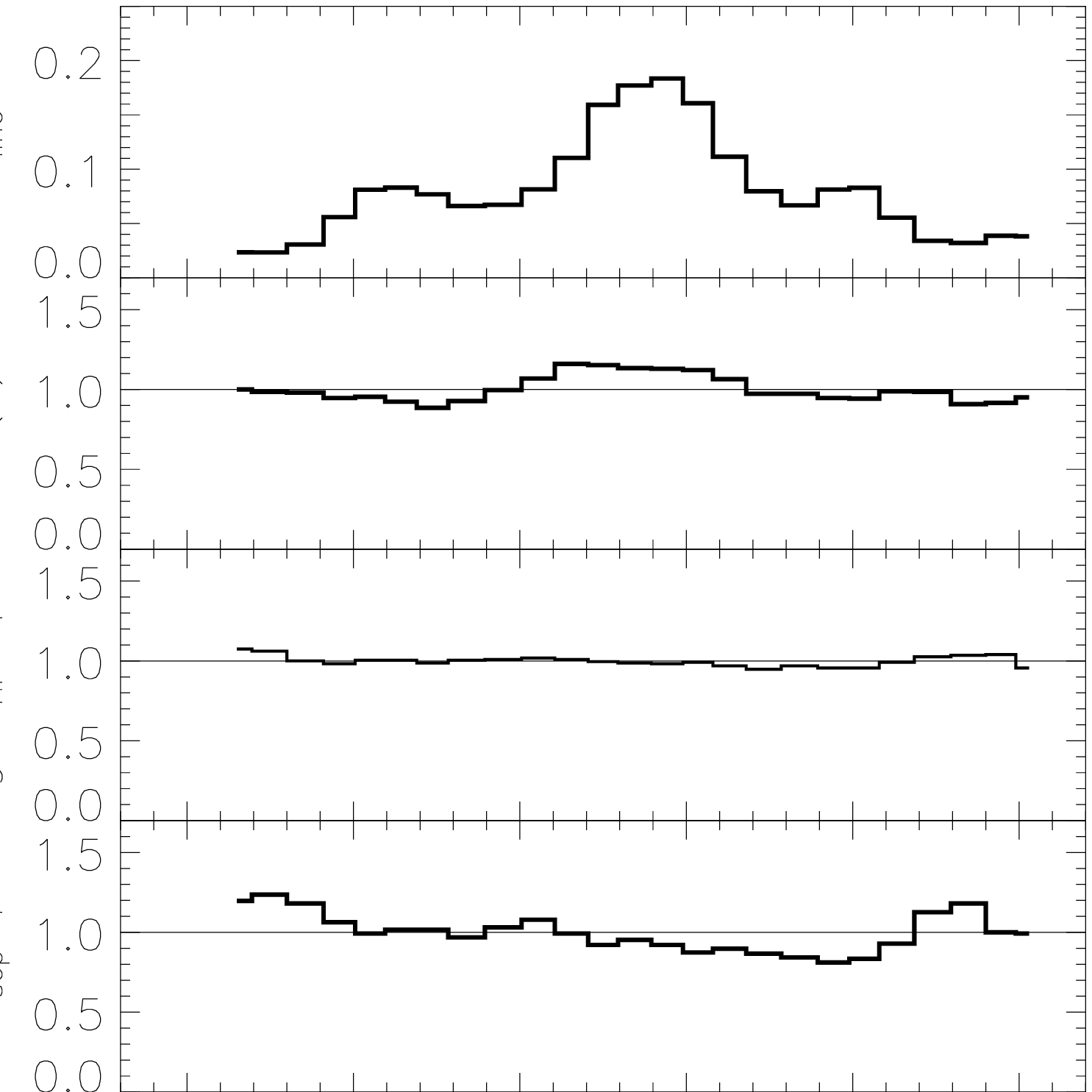}
\vspace{1.cm}
\caption{Redshift variations of fraction of
measured absorbers, $f_{line}=\Delta N_{abs}(z)/N_{abs}$, (top
panel), mean Doppler parameter, $\langle b(z)\rangle$ and HI
column density, $\langle \lg (N_{HI}z_4^{-2})\rangle$ (two
middle panels), and the mean absorber separation, $\langle
d_{sep}z_4^2\rangle$ (bottom panel). All functions are
normalized over the mean values given by (\ref{mnsbhs}).
}
\label{mnbhs}
\end{figure}

For the sample of 6\,270 absorbers the redshift variations of
the three mean observed characteristics of absorbers, namely,
the Doppler parameter, $\langle b\rangle$, the column density of
neutral hydrogen, $\langle\lg N_{HI}^*\rangle=\langle\lg
(N_{HI}/z_4^2)\rangle$, and the mean separation of absorbers,
$\langle d_{sep}^*\rangle=\langle d_{sep}z_4^2\rangle$,
$z_4=(1+z)/4$ are plotted in Fig. \ref{mnbhs} for 1.6 $\leq
z\leq$ 4. These variations are well fitted by
\[
\langle \lg N_{HI}^*\rangle=\langle\lg (N_{HI}/z_4^2)
\rangle= 13.3\pm 0.08,
\]
\be
\langle b\rangle = (26\pm 2.1){\rm km/s},
\quad z_4=(1+z)/4\,,
\label{mnsbhs}
\ee
\[
\langle d_v^*\rangle = \langle 2b/H(z)z_4^{3/2}\rangle=
(0.12\pm 0.01)h^{-1}{\rm Mpc}\,,
\]
\[
\langle d_{sep}^*\rangle = \langle d_{sep}z_4^2\rangle =
(1.3\pm 0.15)h^{-1}{\rm Mpc}\,,
\]
respectively. For the sample of 7\,411 separations we get
\[
\langle d_{sep}^*\rangle = (1\pm 0.1)h^{-1}{\rm Mpc}\,.
\]

Detailed discussion of the observed characteristics of absorbers
can be found, for example, in Kim, Cristiani \& D'Odorico
(2002), Kim et al. (2002, 2004).

\begin{figure}
\centering 
\epsfxsize=7.cm
\epsfbox{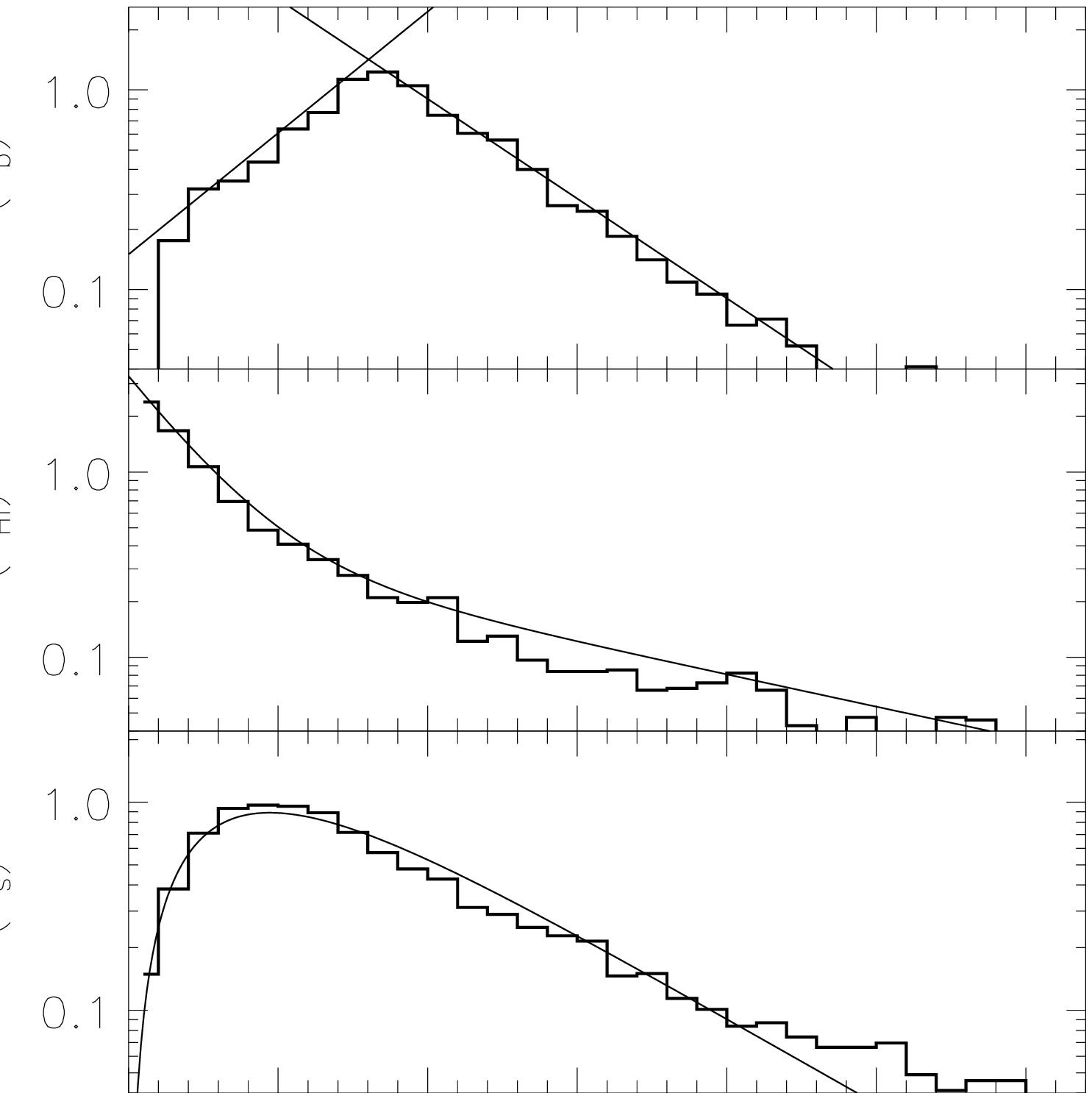}
\vspace{1.cm} 
\caption{Observed PDFs for the Doppler parameter,
$P(x_b)$, the hydrogen column density, $P(x_{HI})$,
and absorbers separations, $P(x_s)$, are plotted together
with fits (\ref{rnbhs}) (solid lines).
}
\label{nbhs}
\end{figure}

The observed probability distribution functions, PDFs, for the
Doppler parameter, $P(b)$, the hydrogen column density,
$P(N_{HI}/z_4^2)$, and the absorbers separation,
$P(d_{sep}z_4^2)$, are plotted in Fig. \ref{nbhs}. Such choice
of variables allows us to suppress the redshift evolution of
the mean characteristics of absorbers.
PDFs for so corrected parameters only weakly vary
with redshift and, as was discussed in Paper I, for the main
fraction of absorbers, these variations do not exceed 10 --
15\%\,. These PDFs are well fitted by exponential functions
\[
P_{fit}(x_{HI})\approx 0.4\exp(-0.8 x_H)+2.9\exp(-5 x_H)\,,
\]
\be
P_{fit}(x_b)\approx \left\{
\begin{array}{cc}
0.15\exp(2.8x_b),& b\leq b_{rap},\cr 0.9\exp(-2.3x_b),& b\geq
b_{rap},\cr
\end{array}\right.
\label{rnbhs}
\ee
\[
P_{fit}(x_s)\approx 3.5\exp(-1.8x_s){\rm erf}^4(\sqrt{1.8x_s})/
\sqrt{x_s}\,,
\]
\[
 x_b={b\over\langle b\rangle},\quad x_{HI}={N_{HI}/z_4^2\over
\langle N_{HI}/z_4^2\rangle},\quad x_s={d_{sep}z_4^2\over\langle
d_{sep}z_4^2\rangle}={d_{sep}^*\over\langle d_{sep}^*\rangle}\,,
\]
where again $z_4=(1+z)/4$ and $b_{rap}\approx 23.5 km/s \sim
\langle b_{bg}\rangle$ (\ref{bg}) discriminates between
absorbers situated in the increasing and decreasing parts of the
PDF $P(x_b)$ in Fig. \ref{nbhs}. The similarity of $b_{rap}$ and
$\langle b_{bg}\rangle$ is an independent confirmation of
estimates (\ref{bg}). For the scatter of measured PDF, $P(x_s)$,
around the fit (\ref{rnbhs}) we have \be
P_{fit}(x_s)/P(x_s)\approx 1.04\pm 0.15,\quad x_s\leq 2\,.
\label{sep_fit} \ee

Any model of the forest has problems explaining the complex
shape of the observed PDF $P(x_b)$ and the existence of
absorbers with $b\leq b_{bg}$. These absorbers are sometimes
related to the unidentified metal lines what is perhaps possible
for colder absorbers. However, large fraction of such absorbers
($\sim 40\%$) indicates that majority of them must be related to
usual hydrogen clouds formed within regions with lower
background temperature. Our analysis shows that statistical
characteristics of these absorbers are consistent with expected
ones for absorbers formed due to adiabatic compression in
rapidly expanded regions (see Secs. 3.5 \& 4.5). Examples of
such absorbers were also found in simulations (see, e.g.,  Bi \&
Davidsen 1997; Zhang et al. 1998; Dav\'e et al. 1999). This
subpopulation can also contain some number of "artificial"
caustics (McGill 1990)\,.

The mean size of rapidly expanded
regions, $\langle D_{rap}\rangle$, can be measured as a
distance along the line of sight between two absorbers with
$b\geq b_{rap}$ closest to a colder absorber. It increases
with time as
\be
\langle D_{rap}z_4^2\rangle\approx 1.6\pm
0.2 h^{-1}{\rm Mpc}\,,
\label{creg}
\ee
and the typical mass associated with these regions is
\be
M_{rap}\sim {\pi\over 6}\langle\rho(z)\delta_m D_{rap}^3
\rangle\sim 10^{13}z_4^{-3}\langle\delta_m\rangle M_\odot
\left({ \Omega_m\over 0.3}{0.7\over h}\right)\,,
\label{tmass}
\ee
where $\delta_m=\rho_m/\langle\rho_m\rangle\leq 1$\,. At
$z_4\sim$ 1 the mass
$M_{rap}$ is in the range of galactic masses and it rapidly
increases with time. This result is consistent with the expected
symmetry of positive and negative initial density perturbations
what leads to formation of both galaxies and rapidly expanded
regions.

Even these results allow one to obtain some inferences on
the evolution of the forest:
\begin{enumerate}
\item{}Regular redshift variations of the mean observed
characteristics of absorbers (\ref{mnsbhs}) and weak redshift
dependence of PDFs (\ref{rnbhs}) indicate the self similar
character of absorber's evolution over the whole range of
redshifts under consideration. So, we can conclude, that at
these redshifts the evolution is dominated by a balanced action
of the same physical factors.
\item{}The complex form of the
PDF $P(b)$ confirms that  absorbers have been  formed within
both rapidly and moderately expanded regions. Analysis of these
subpopulations of absorbers allows one to estimate some
parameters of such regions. We will discuss this problem
 in Secs. 3.5\,\&\,4.5\,.
\item{}The very wide range of measured Doppler parameters,
0.2 $\leq b/b_{bg}\leq$ 5, indicates a wide variety of
initial perturbations. Such a wide range of Doppler
parameters is not reproduced in simulations (Meiksin et al.
2001).
\item{} The decline of the measured $\langle N_{HI}\rangle
\propto (1+z)^2$ could be related to the retained expansion of
majority of observed absorbers in the transverse directions.
\item{}The growth of the mean observed separation between absorbers,
$\langle d_{sep}\rangle\propto (1+z)^{-2}$ indicates the progressive
decline of the number density of observed absorbers that can be
related to the decrease of their hydrogen column density under the
observational limit $N_{HI} \approx 10^{12}cm^{-2}$.
\item{}Comparison of measured $\langle d_v(z)\rangle$ and
$(1+z)^{-1}\langle d_{sep}(z)\rangle$ (\ref{mnsbhs}) indicates
that at $z\geq 4$ the expected overlapping of absorbers becomes
essential and, perhaps, absence of observed absorbers in spectra
of the farthest QSO (Fan et al. 2002, 2003, 2004) can be partly
related to this effect.
\end{enumerate}

\section{Model of absorbers formation and evolution}

\subsection{Physical model of absorbers}

Many models were proposed during the last twenty years to
explain regular redshift variations of the mean observed
characteristics of absorbers (\ref{mnsbhs}) and weak redshift
dependence of PDFs (\ref{rnbhs}) (see references in Rauch
(1998), and in Paper I and Paper II). The simplest one connects
the absorber characteristics with a suitable set of early formed
equilibrium clouds. This model naturally explains the weak
redshift dependence of the mean Doppler parameter, $\langle
b\rangle$, and the observed PDFs (\ref{rnbhs}). It explains also
the regular redshift variations of the mean absorber separation
(\ref{mnsbhs}). Indeed, the mean proper free path between
absorbers is \be (1+z)^{-1}\langle d_{sep}(z)\rangle\propto
(1+z)^{-3}\propto \langle n_{abs}(z)S_{abs}(z)\rangle^{-1}\,,
\label{freepath} \ee where $n_{abs}\,\&\,S_{abs}$ are the number
density and the surface area of such clouds orthogonal to the
line of sight. In this case we have from (\ref{freepath})
\[
n_{abs}(z)\propto (1+z)^3,\quad S_{abs}(z)\propto const(z)\,.
\]

However, it is well known that the existence of such low density
equilibrium clouds cannot be reasonably explained. Moreover, with
this model the observed evolution of the mean column density of
neutral hydrogen, $\langle N_{HI}\rangle\propto (1+z)^2$, must be
related to the monotonic evolution of the UV background,
$\langle\Gamma_{12}\rangle\propto (1+z)^{-2}$, what disagrees with
the available estimates of $\Gamma_{12}$.

Numerous simulations indicate that the majority of the LSS elements
can be related to the extended anisotropic moderate density clouds
with a complex internal structure and low density envelope. Such
long lived clouds are relaxed along the shorter axis and are
expanded/compressed along transverse directions. The formation of DM
pancakes as an inevitable first step of evolution of small
perturbations was firmly established both by theoretical
considerations (Zel'dovich 1970; Shandarin \& Zel'dovich 1989) and
numerical simulations (Shandarin et al. 1995). The anisotropic
galaxy walls such as the Great Wall are observed in the SDSS and 2dF
galaxy surveys and represent examples of the Zel'dovich pancakes.
This suggests that absorbers can also be linked with  the Zel'dovich
pancakes and the reasonable model of the forest evolution can be
constructed on the basis of the Zel'dovich theory.

However, in some papers devoted to the hydrodynamical simulations of
the forest evolution absorbers are often identified with unrelaxed
moderate density clouds and their Doppler parameter $b$ is related
to the gradient of velocity of infalling matter rather than to
thermal velocities. Of course, observed samples include some
fraction of such objects mainly among weaker absorbers. However,
properties of richer absorbers with metal systems are found to 
be in a good
agreement with a model assuming local hydrostatic equilibrium (see,
e.g. Carswell, Schaye \& Kim 2002; Telferet al. 2002; Simcoe,
Sargent, Raugh 2002, 2004; Boksenberg, Sargent, Raugh 2003; Manning
2002, 2003 a,b; Bergeron \& Herbert-Fort 2005). This means that,
possibly, the fraction of simulated extended unrelaxed clouds is
artificially enhanced by technical limitations (see, e.g.,
discussion in Meiksin, Bryan \& Machacek 2001; Manning 2003 and
Paper II). It is important that these simulations cannot yet
reproduce well enough the characteristics of observed absorbers
discussed in previous Section.

This discussion shows that an adequate physical model of the
complex evolution of the forest has not been proposed yet.
Indeed, the relaxation of DM pancakes leads to complex internal
structure of pancakes, the adiabatic compression and/or
expansion of absorbers in transverse directions is changing
their overdensity and temperature, the radiative cooling and
bulk heating leads to the drift of the gas entropy and
overdensity but leaves unchanged the depth of the potential well
formed by DM distribution. Merging of pancakes increases more
strongly the depth of the potential well and the gas entropy but
the overdensity of the gaseous component increases only
moderately. The temperature and overdensity of the trapped gas
are rearranged in accordance with the condition of hydrodynamic
equilibrium across the pancake continuously. These processes
imply the existence of a complicated time-dependent internal
structure of absorbers.

In this paper we discuss the model of absorbers evolution based
on the Zel'dvich theory. This analytical model links the self
similar evolution of dominated DM component with observed
characteristics of absorbers. Here we assume that:
\begin{enumerate}
\item{} The DM distribution forms a set of sheet--like clouds
(Zel'dovich pancakes), their basic parameters are approximately
described by the Zel'dovich theory of gravitational instability
applied to the CDM or WDM initial power spectrum (DD99; DD04).
The majority of DM pancakes are partly relaxed, long-lived, and
their properties vary due to the successive accretion of matter,
merging and expansion and/or compression in the transverse
directions.
\item{} Gas is trapped in the gravitational potential wells
formed by the DM distribution. For majority of absorbers,
the observed Doppler parameter, $b$, traces the gas
temperature and the depth of the DM potential wells. We
consider the possible macroscopic motions within pancakes
as subsonic and assume that they cannot essentially distort
the measured Doppler parameter.
\item{} The gas is ionized by the UV background and for the
majority of absorbers ionization equilibrium is assumed.
\item{} For a given temperature, the gas density within the
potential wells is determined by the gas entropy created during
the previous evolution. The gas entropy is changing, mainly, due
to relaxation of the compressed matter, shock heating in the
course of merging of pancakes, bulk heating by the UV background
and local sources and due to radiative cooling. These processes
slowly change the entropy and density of the trapped gas. Random
variations of the intensity and spectrum of the UV background
enhance random scatter of the observed properties of absorbers.
\end{enumerate}

In this model we expect that the mean surface density of both DM and
baryonic components of pancakes weakly varies with time and the mean
density of matter compressed within absorbers decreases as
$\langle\rho_{abs}\rangle\propto (1+z)^\upsilon,\,\upsilon\sim 1.5 -
1$ for colder and hotter absorbers, respectively. We cannot
discriminate between evolution of the absorbers number density,
$n_{abs}$, and their surface area, $S_{abs}$. However, numerical
simulations demonstrate that the general tendency of evolution is a
sequential growth of masses and sizes of pancakes accompanied by 
a drift of weaker ones under the observational limit and decrease of
the comoving number density of observed absorbers. Thus, as is seen
from (\ref{freepath}), if the surface area of absorbers increases
$\langle S_{abs}\rangle\propto (1+z)^{-\kappa}$ then the number
density of observed absorbers decreases as $\langle n_{abs}
\rangle\propto (1+z)^{3+\kappa}$. We show that the observed
evolution of mean hydrogen column density and mean absorber
separations (\ref{mnsbhs}\,\&\,\ref{rnbhs}) coincides with
theoretical expectations for Gaussian initial perturbations. 

The wide range of the observed Doppler parameters, 5--10 km/s
$\leq b\leq$ 100 km/s, demonstrates a complex composition of the
forest. In this paper we roughly identify three subpopulations
of absorbers, namely, hot absorbers formed in the course of
merging and shock compression, warm absorbers formed due to
adiabatic and weak shock compression in moderately expanded
regions, and an unexpectedly rich subpopulation of colder
absorbers. We link the measured Doppler parameter with the depth
of 1D potential well formed by the compressed DM component and
we neglect the contribution of macroscopic velocities what
restricts the achieved precision of our approach.  Non the less
this model reproduces quite well the self similar evolution of
absorbers and allows one to reconstruct the initial correlation
function of velocities and the initial power spectrum down to
very small scales.

\subsection{The initial power spectrum and correlation functions
of the initial velocity field}

In this Section we summarize the main results obtained in DD99 and
DD04 concerning the evolution of DM pancakes and in Sec. 3.9 we
show how to improve the estimates of the correlation function of the
initial velocity field, and the shape of initial power spectrum
discussed in Paper II.

As a reference power spectrum of initial perturbations we take
the standard CDM--like spectrum with the Harrison -- Zel'dovich
asymptotic, \be P(k)={A^2k\over k_0^4}T^2(\eta)D_W(\eta),\quad
\eta={k \over k_0},\quad k_0={\Omega_mh^2\over{\rm Mpc}}\,,
\label{f1} \ee where $A$ is the dimensionless amplitude of
perturbations, $k$ is the comoving wave number. The transfer
function, $T(\eta)$, and the damping factor, $D_W(\eta)$,
describing the free streaming of DM particles were given in
Bardeen et al. (1986). For WDM particles the dimensionless
damping scale, $R_f$, and the damping factor, $D_W$, are
\[
R_f={1\over 5}\left({\Omega_mh^2{\rm keV}}\over M_{DM}
\right)^{4/3},\quad D_W=\exp[-\eta R_f-(\eta R_f)^2]\,,
\]
where $M_{DM}$ is the mass of WDM particles in keV (Bardeen et
al. 1986). This relation illustrates the clear dependence of
characteristics of small scale perturbations on the mass of DM
particles.

For the spectrum (\ref{f1}) the coherent lengths of velocity and
density fields, $l_v\,\&\,l_\rho$, are expressed through the
spectral moments, $m_{-2}\,\&\,m_0$, (DD99) as follows: \be
l_v={1\over k_0\sqrt{m_{-2}}}={6.6\over \Omega_mh^2} {\rm
Mpc}\approx 31.4 h^{-1}{0.21\over \Omega_mh}{\rm Mpc} \,,
\label{lv} \ee
\[
m_{-2} = \int_0^\infty d\eta~\eta T^2(\eta)D_W(\eta)
\approx 0.023,\quad l_\rho=q_0 l_v\,,
\]
\[
m_0 = \int_0^\infty d\eta~\eta^3 T^2(\eta)D_W(\eta),
\qquad q_0=5{m_{-2}^2\over m_0}\,.
\]
The moment $m_0$ depends upon the mass of dominant component of
DM particles and as is shown in Sec. 4.2, $q_0\leq 10^{-3}$,
$m_0\geq 2.4$, $M_{DM}\geq$ 1\,MeV\,.

As was demonstrated in DD99 and DD04, the basic statistical
characteristics of structure are expressed through the
normalized longitudinal correlation function of the initial
velocity field, $\bf v(\tilde{\bf q})$\,,
\be
\xi_v(l_vq)=3{\langle({\bf q\cdot v}(\tilde{\bf q}_1)) ({\bf
q\cdot v}(\tilde{\bf q}_2))\rangle\over \sigma_v^2q^2},\quad{\bf
q}={\tilde{\bf q}_1-\tilde{\bf q}_2\over l_v}\,.
\label{xiv0}
\ee
Here $\tilde{\bf q}_1 ~\&~ \tilde{\bf q}_2$ are the
unperturbed coordinates of two particles at $z=$ 0, $q=|{\bf
q}|$, and $\sigma_v^2$ is the velocity variance. This function
is expressed through the power spectrum: \be \xi_v(l_vq)={3\over
m_{-2}}\int_0^\infty d\eta\,\eta^2\cos x \int_\eta^\infty
{dy\over y^2}T^2(y)D_W(y)\,, \label{xivv} \ee
\[
\eta T^2(\eta)D_W(\eta)={\sqrt{m_{-2}}\over 3}\int_0^\infty
(2\cos x+x\sin x)\xi_v(l_vq)dq\,,
\]
\[
x=kl_vq,\quad \eta=k/k_0,\quad \xi_v(0)=1,\quad \int_0^\infty
dq~\xi(l_vq)=0\,.
\]
Similar relations can be also written for any initial power
spectrum $P(k)$\,.

For the CDM -- like spectrum (\ref{f1})
with $q_0< 10^{-3}$ and for the most interesting
range $0.5\geq q$ the velocity
correlation function can be fitted as follows:
\be
\xi_v(q)=\xi_{CDM}\approx 1-{1.5 q^2\over\sqrt{2.25q^4+q^2+
p_0^{1.4}q^{0.6}+q_0^2}}\,,
\label{xiv}
\ee
where $p_0\approx 1.1\cdot10^{-2}$ and $q_0$ was introduced
in (\ref{lv}). Further on, we will use this function
as the reference one and will compare it with observational
estimates of $\xi_v(q)$.

As was shown in DD99 and DD04, the main DM characteristics
depend upon the self similar variable
\be
\zeta(q,z) ={q^2\over 4\tau^2(z)[1-\xi_v(q)]}\,,
\label{xit}
\ee
where the 'time' $\tau(z)$ describes the growth of perturbations
due to the gravitational instability. For the $\Lambda$CDM
cosmological model (\ref{basic}), and for $z\geq$ 2 we have
\be
\tau(z)\approx \tau_0\left({1+1.2\Omega_m\over 2.2\Omega_m}
\right)^{1/3}{1\over 1+z}\approx {1.27\tau_0\over 1+z}\,,
\label{Btau}
\ee
and $\tau_0$ characterizes the amplitude of initial
perturbations. It is proportional to $\sigma_8$, the
variance of the mass within a randomly placed sphere of
radius 8$h^{-1}$Mpc. Latest estimates (Spergel et al. 2003, 2006;
Viel et al. 2004b) for the model (\ref{basic}) are
\be
\sigma_8\approx 0.9\pm 0.1,\quad \tau_0=0.21\sigma_8\approx
(0.19\pm 0.02){\sigma_8\over 0.9}\,.
\label{tau0}
\ee

\subsection{Expected characteristics of DM absorbers}

In this section we introduce (without proofs) the basic
characteristics of DM pancakes as a basis for further analysis.
For more details see DD99 and  DD04.

\subsubsection{DM column density of absorbers}

The fundamental characteristic of DM pancakes is the
dimensional, $\mu$, or the dimensionless, $q$, Lagrangian
thickness (the dimensionless DM column density) :
\be
\mu\approx {\langle\rho_m(z)\rangle l_vq\over (1+z)} =
{3H_0^2\over 8\pi G}l_v\Omega_m(1+z)^2q\,,
\label{mu}
\ee
where $l_v$ is the coherent length of initial velocity field
(\ref{lv}). The Lagrangian thickness of a pancake, $l_v q$,
is defined as the unperturbed distance at redshift $z=0$
between  DM particles bounding the pancake (\ref{xiv0}).

As was found in DD99 and DD04 for Gaussian initial
perturbations, the expected probability distribution
function for the DM column density is
\be
N_q(\zeta)\approx {2\over\sqrt{\pi}} e^{-\zeta}~{{\rm erf}
(\sqrt{\zeta})\over\sqrt{\zeta}},\quad W_q(<\zeta)
={\rm erf}^2(\zeta)\,,
\label{qpdf}
\ee
\[
\langle\zeta\rangle\approx{1\over 2}+{1\over\pi}\approx 0.82,
\quad \langle q\rangle\approx 6\tau^2\langle\zeta\rangle\approx
{2\pm 0.2\over z_4^2} 10^{-2}\left({\tau_0\over 0.2}
\right)^2\,,
\]
where $z_4=(1+z)/4$\,, $\zeta(q,z)$ was introduced in (\ref{xit})
and $W_q(<\zeta)$ is the cumulative probability function.

Strictly speaking, the relations (\ref{mu}) and (\ref{qpdf}) are
valid for pancakes formed and observed at the same redshift
$z_{obs}=z_f$ because after pancake formation the transverse
expansion and compression changes its DM column density and
other characteristics. However, owing to the symmetry of
moderate distortions of general expansion in transverse
directions, these processes do not change the statistical
characteristics for majority of pancakes observed at redshift
$z_{obs}\leq z_f$ (DD04). This means that statistically we can
consider each pancake as created at the observed redshift.
However, the symmetry is distorted for rapidly expanded regions
observed as absorbers with $b\ll b_{bg}$ because the
cross--section of strongly compressed absorbers is small and,
therefore, they are rarely observed. More details are given in
Secs. 3.5, 3.6 and in DD04.

\subsubsection{Proper sizes of absorbers}

The actual thickness of a DM pancake is estimated as
\be
d_{abs} = {\mu\over\rho_m} ={l_vq\over (1+z)\delta_m},
\quad \delta_m=\rho_m/\langle\rho_m(z)\rangle\,.
\label{delrm}
\ee
Here $\delta_m$ is the mean overdensity of compressed matter
above the background density.

 For the transverse size of absorbers, the expected
characteristics were estimated in DD04 as follows:
\be
N_{tr}(\zeta_{tr})\approx {2\over\sqrt{\pi}}\exp(-\zeta_{tr}^2),
\quad \zeta_{tr}^2\approx {q_{tr}\over 6\tau^2(z)}\,,
\label{trsize}
\ee
\[
\langle \zeta_{tr}^2\rangle\approx 1/2,\quad
\langle l_vq_{tr}\rangle\approx 3\tau^2(z)l_v\approx
0.45z_4^{-2}h^{-1}{\rm Mpc}\,.
\]
Here again $z_4=(1+z)/4$, $z$ is the observed redshift and
$l_vq_{tr}$ is the expected size of absorbers at $z=0$.

\subsubsection{Absorbers separation}

An important characteristic of the distribution of absorbers is
their separation determined as the distance along the line of
sight between centers of neighboring absorbers observed with the
separation $\Delta z$, \be d_{sep}={c\Delta z\over
H(z)}=5.5\cdot 10^3{\Delta z\over (1+z)^{3/2}
}\sqrt{0.3\over\Omega_m}h^{-1}{\rm Mpc}\,. \label{sep} \ee This
separation is identical to the free path between absorbers and
is quite similar to the popular descriptor \be {\Delta
N_{abs}\over\Delta z}\propto \langle d_{sep}(z) \rangle^{-1}\,.
\label{dndz} \ee

For Gaussian initial perturbations the PDF for the absorber
separations in Lagrangian space, $d_L$, is
\[
N_L(\zeta_L)\approx 4\pi^{-1/2}\exp(-\zeta_L)[(1+2\zeta_L)
D_w(\sqrt{\zeta_L})-\sqrt{\zeta_L}]
\]
\be
\approx 2.82\exp(-\zeta_L){\rm erf}^4(\zeta_L)/\sqrt{\zeta_L}\,,
\label{Lsep}
\ee
\[
\zeta_L = \zeta(q_L,z),\quad q_L=d_L/l_v,\quad\langle\zeta_L
\rangle\approx 1.5\,,
\]
where $\zeta(q,z)$ was introduced by (\ref{xit}) and
\[
D_w(x)=\int_0^x dy\exp(y^2-x^2)
\]
 is the Dawson function.

Comparing the pancake surface density, $\mu(z)$, with the
absorbers separation, $d_{sep}$, we can also estimate the
fraction of matter accumulated by absorbers as
\be
f_{abs}\simeq{\mu(1+z)\over\langle\rho_m\rangle d_{sep}}
={l_vq\over d_{sep}},\quad \langle f_{abs}\rangle\approx
{\langle q\rangle\over\langle q_L\rangle}\approx 0.55\,.
\label{fract}
\ee
However, this formal estimate is of limited
significance because it does not consider the complex processes
of absorbers evolution. More detailed 3D analysis (DD04) shows
that at a later period of absorbers evolution we can expect \be
f_{abs}\simeq 0.38 - 0.44 \label{ffr} \ee

The observed characteristics of absorbers are measured in the
redshift space where both the proper motions of absorbers and
their peculiar velocities distort the PDF (\ref{Lsep}) and, in
particular,  lead to the merging and artificial blending of
absorbers. These processes are driven by the spatial modulations
of gravitational potential formed by the large scale
perturbations and can be also described in the framework of
Zel'dovich approximation (DD99). However, such description
depends upon the size of absorbers (\ref{delrm})
which variations with redshift cannot be described
theoretically. So, characteristics of absorbers in both real and
redshift spaces can be determined only approximately.

For the PDF of absorbers separation in the redshift space we get \be
N_{sep}\approx2.4\exp(-1.35x_{rd})[1-0.85\exp(-1.35x_{rd})]
\label{sep_mns}
\ee
\[
x_{rd}={\zeta(q_{rd},z)\over\langle \zeta(q_{rd},z)\rangle},
\quad \langle\zeta(q_{rd},z)\rangle
\approx 2.13,\quad q_{rd}={d_{sep}\over l_v}\,,
\]
\[
\langle d_{sep}\rangle
\approx 6l_v\tau^2\langle\zeta(q_{rd},z)\rangle\approx 1.3z_4^{-2}
h^{-1}{\rm Mpc}(\tau_0/0.2)^2,
\]
where $z_4=(1+z)/4$ and $\zeta(q,z)\,\&\,d_{sep}$ are defined by
(\ref{xit}) and (\ref{sep}). These expressions describe correctly
properties of larger separations but become unreliable for smaller
separations where the influence of the proper sizes of absorbers is
more important. In the real space the expression (\ref{sep_mns})
also approximates the PDF of distances between neighboring absorbers
but the mean value $\langle \zeta_{real}\rangle\approx$ 1.64 is
smaller than that in the redshift space.

The same approach allows us to determine the expected
characteristics of merged absorbers. Instead of (\ref{qpdf}), in
redshift space, we get for such absorbers \be N_{mrg}(x)\approx
1.3\exp(-1.1x)[1-0.5\exp(-2.2x)]\,, \label{merg} \ee
\[
x=\zeta_{mrg}/\langle\zeta_{mrg}\rangle,\quad
\zeta_{mrg}=\zeta(q_{mrg},z),\quad
\langle\zeta_{mrg}\rangle=1.22\,,
\]
where $\zeta(q,z)$ was introduced in (\ref{xit}) and $q_{mrg}$
is the dimensionless DM column density of merged absorbers. This
relation indicates that the population of poorer merged
absorbers is suppressed and, in particular, the mean DM column
density of merged absorbers is larger than that for all
absorbers as given by (\ref{qpdf}).

The relations (\ref{qpdf}), (\ref{fract}), (\ref{sep_mns}),
(\ref{merg}) show that during the self similar period of
structure evolution, when the relations (\ref{xiv}) and
(\ref{Btau}) are valid, we can expect regular variations of the
basic characteristics of absorbers such as their DM column
density, $q$, separation, $d_{sep}$, and fraction of matter
accumulated by absorbers, $f_{abs}$. These regular variations
are distorted at small redshifts when the growth of
perturbations is decelerated and at higher redshifts, when
the blending of absorbers becomes more important.

\subsection{Doppler parameters of absorbers}

For relaxed and gravitationally confined absorbers their Doppler
parameters are closely linked to the potential wells formed by
the DM distribution. As is well known, for an equilibrium slab
of DM the depth of its potential well is
\be
\Delta \Phi\approx
{\pi G\mu^2\over \langle \rho(z)\rangle \delta_m}\Theta_\Phi =
{3\over 8}v_0^2{q^2(1+z) \over\delta_m}\Theta_\Phi\,,
\label{phi}
\ee
\[
v_0=H_0l_v\sqrt{\Omega_m}=1\,720 km/s\sqrt{0.15\over
\Omega_mh^2}\,.
\]
where the random factor $\Theta_\Phi$
characterizes the inhomogeneity of DM distribution across the
slab and the evaporation of matter in the course of its
relaxation.

Analysis of numerical simulations (Demia\'nski et al. 2000)
indicates that the relaxed distribution of DM component can be
approximately described by the polytropic equation of state with
the power index $\gamma_m\approx$ 1.5 - 2\,. Thus, for
$\gamma_m=2$ the equilibrium density profile across the slab can
be directly found and
\[
\Theta_\Phi= 4/\pi\approx 1.3 \,.
\]
The actual distribution of DM component across a slab and the
value of $\Theta_\Phi$ depends upon the relaxation process which
is essentially accelerated by the process of pancake disruption
into the system of high density clouds and filaments. In the
course of relaxation $\sim$ 10 -- 15\% of matter is evaporated
what decreases the factor $\Theta_\Phi$. This means that the
expected $\Theta_\Phi\sim 1$ randomly varies from absorber to
absorber.

The Doppler parameter is defined by the depth of potential well
(\ref{phi}) and for the isentropic gas with $p_{gas}
\propto\rho_{gas}^{5/3}$ trapped within the well, we get
\be
b^2\approx b_{bg}^2+{4\over 5}\Delta \Phi\approx b_{bg}^2
+{3\over 10}v_0^2{q^2\over\delta_m}(1+z)\Theta_\Phi\,.
\label{bb}
\ee
Variations of the gas entropy across absorbers increase the
random variations of $\Delta\Phi\,\&\,b$.

For hot absorbers with $b\gg\langle b_{bg}\rangle$ we can
neglect the difference between the actual and mean background
temperature and in (\ref{bb}) use $\langle b_{bg}\rangle$
instead of $b_{bg}$. In such a way we link $q^2/\delta_m$ with
$b\, \&\,\langle b_{bg}\rangle$ with a reasonable precision.

\subsection{Absorbers within rapidly expanded regions}

For significant fraction of absorbers -- up to 20\% -- the Doppler
parameter is smaller then the expected mean background one
(\ref{bg}). Such absorbers are often related to unrecognized metal
lines. However, both theoretical arguments and numerical simulations
show that such absorbers can also be related to hydrogen clouds
formed within colder rapidly expanded regions.

The temperature of relaxed HI absorbers formed by the compression of
matter (\ref{bb}) cannot be smaller than the background one
(\ref{bg}). However, within low density rapidly expanded regions the
background temperature given by (\ref{rap}) is smaller than the mean
one, and in such regions the hydrogen clouds with $b_{bg}\leq b
\leq\langle b_{bg}\rangle$  can be formed. Our analysis (Sec. 4.5)
shows that majority of colder absorbers could be related to such
hydrogen clouds.

According to the Zel'dovich theory of gravitational instability and
for Gaussian initial perturbations, regions with moderate
distortions of the cosmological expansion dominate and probability
to find rapidly expanded or compressed regions is exponentially
small. However, the observations of galaxies at redshifts under
consideration corroborate the existence of strong distortions of
cosmological expansion at least on galactic scales. This means that
owing to the symmetry of positive and negative initial density
perturbations we should also observe rapidly expanded regions with
small background density and temperature. Number of such regions
exponentially decreases for larger distortions of the expansion.

Owing to the same symmetry, for majority of absorbers, fraction of
absorbers with a moderate random expansion and compression in the
transverse directions are close to each other and the influence of
this factor weakly distorts the mean absorbers characteristics.
However, within rapidly expanded regions {\it all} absorbers are
adiabatically expanded in the transverse directions what, in
particular, decreases the fraction of neutral hydrogen,
\[
x_H\propto n_b/b^{3/2}\Gamma_{12}\propto n_b^{1/2}/\Gamma_{12}\,,
\]
and its observed column density, and leads to a systematic drift
of absorbers under the observational limit $lg N_{HI}\sim 12$.

Theoretical estimates show that owing to the correlation of
velocity perturbations across the pancake and in the transverse
directions the rate of pancakes formation within rapidly
expanded regions is smaller than the mean one. For the CDM like
initial power spectrum the coefficient of correlation of
orthogonal velocities is $c_v\approx 1/3$ (DD04). In this case,
for the fraction of matter, $f_{rap}$, and the mean column
density of DM component, $q_{rap}$ in the rapidly expanded
regions, we expect
\be
f_{rap}\approx 0.33\langle f_{abs}\rangle, \quad\langle\zeta_{rap}
\rangle=\langle\zeta(q_{rap},z)\rangle\approx 0.5\langle\zeta
\rangle\,.
\label{nrap}
\ee

\subsection{Characteristics of the gaseous component}

The observed column density of the neutral hydrogen can be
written as an integral over the line of sight through a pancake
\be N_{HI} = \int dx ~\rho_b x_H = 2\langle x_H\rangle{\langle
n_b(z)\rangle l_v q\over 1+z}{0.5\over cos\theta}\,. \label{NH1}
\ee Here $\langle x_H\rangle$ is the mean fraction of the
neutral hydrogen and $cos\theta$ takes into account the random
orientation of absorbers and the line of sight ($\langle
cos\theta\rangle\approx 0.5)$. As was noted in Sec. 3.4, we
assume also that both DM and gaseous components are compressed
together and, so, the column densities of baryons and DM
component are proportional to each other.

Under the  assumption of ionization equilibrium of the
gas (\ref{ieq}) and neglecting a possible contribution
of macroscopic motions to the $b$-parameter ($T\propto
b^2$), for the fraction of neutral hydrogen and its
column  density we get:
\[
\langle x_H\rangle = x_0\delta_b\beta^{-3/2}
(1+z)^{33/14}\Theta_x,\quad \beta=b/\langle b_{bg}\rangle\,,
\]
\be
{N_{HI}\over N_0} = {q\delta_b\over\Gamma_{12}\beta^{3/2}}
(1+z)^{61/14},\quad N_0 = 5.5\cdot 10^{12}cm^{-2}
\Theta_H,
\label{NH2}
\ee
\[
\Theta_H={\Theta_x\over\Theta_{bg}^{3/7}}{0.15\over
\Omega_mh^2}{\langle cos\theta\rangle\over cos\theta}
\left({\Omega_bh^2\over 0.02}\right)^2,\quad\delta_b=
n_b/\langle n_b\rangle\,,
\]
where $\Gamma_{12}$, $\langle b_{bg}\rangle$, $\Theta_{bg}$
and $x_0$ were defined in (\ref{bg}) and (\ref{xbg}) and the
factor $\Theta_x\sim 1$ describes the inhomogeneous distribution
of ionized hydrogen along  the line of sight.

However, the overdensity of the baryonic component, $\delta_b$
is not identical to the overdensity
of DM component, $\delta_m$, (see, e.g., discussion in Matarrese
\& Mohayaee 2002). Indeed, the gas temperature and the Doppler
parameter are mainly determined by the characteristics of DM
component (\ref{bb}) but the gas overdensity is smaller than
that of DM component due to larger entropy of the gas. Moreover,
the bulk heating and cooling change the density and entropy of
the gas trapped within the DM potential well.  These processes
change the baryonic density of pancakes and we can write
\be
\delta_b = \Theta_b(z)\delta_m,\quad \Theta_b(z)\leq 1\,.
\label{kappab}
\ee
The factor $\Theta_b$ should be small for absorbers formed
due to adiabatic and weak shock compression because of the
large difference between entropies of the background DM and
the gas, and $\Theta_b\rightarrow$ 1 for richer hot absorbers
formed due to strong shock compression when entropies of both
components are comparable.

Similarly to the proper thickness of a DM pancake (\ref{delrm}), the
thickness of a gaseous pancake is estimated as \be
d_{abs}={l_vq\over (1+z)\delta_b}\,. \label{delr} \ee For
adiabatically compressed absorbers it is larger then the thickness
of the DM pancake but for shock compressed absorbers they are close
to each other. So defined $d_{abs}$ can be compared with the
estimated redshift thickness of absorbers determined by the observed
Doppler parameter, \be d_v = 2 b/H(z)\,. \label{br} \ee We can
expect that, as usual, $d_v\geq d_{abs}$ owing to the impact of
thermal velocities. In spite of the limited precision of
determination of $d_v\,\&\,d_{abs}$ these estimates allow us to
correct the derived DM column density, $q$.

For long lived absorbers the influence of the bulk heating can be
estimated in the same manner as it was done for characteristics
of the background (Demia\'nski \& Doroshkevich 2004b). Solving
the equation of thermal balance for absorbers formed at $z=z_f$
and observed at $z\ll z_f$ we obtain for the entropy of
compressed gas:
\be
F_s^{3/2}(z)\approx F_s^{3/2}(z_f)+{8\over
7}{F_{bg}^{3/2} (z)\over\beta^{1/2}(z)}\left[1-\left({1+z\over
1+z_f} \right)^{3/2}\right]\,.
\label{thb}
\ee
As is seen from
this relation, the bulk heating is negligible for shock
compressed absorbers with $F_s(z_f)\gg F_{bg} (z)$, $\beta(z)\gg
1$. For adiabatically compressed long lived absorbers with
$z_f\gg z$, $F_s(z_f)\approx F_{bg}(z_f)\leq F_{bg}(z)$ we get
\be
F_s^{3/2}(z)\approx F_{bg}^{3/2}(z)/\beta^{1/2}(z),\quad
\delta_b\approx \beta^{7/2}\,.
\label{dltth}
\ee
This result
demonstrates that the bulk heating is specially important for
absorbers with $\beta\ll 1$ formed within rapidly expanded
regions. For such absorbers the merging is suppressed and their
number decreases mainly owing to the drift under the
observational limit.

\subsection{Observed characteristics of absorbers}

Eqs. (\ref{bb}) and (\ref{NH2}) relate three independent
variables, namely, $q, \delta_m \,\&\,\delta_b$. To find the DM
column density, $q$, it is therefore necessary to use an
additional relation which connects the basic parameters of
absorbers. Here we assume that the richer absorbers with
$b\geq b_{thr}=\beta_{thr}\langle b_{bg}\rangle\,> \langle
b_{bg}\rangle$ are formed due to shock compression and this
process is accompanied by strong relaxation of compressed
matter. For such absorbers $\delta_b\approx \delta_m$ and
their DM column density is:
\[
\delta_b\approx\delta_m\approx {v_0^2(1+z)\over
b^2-\langle b_{bg}\rangle^2}q^2,\quad  b\geq b_{thr}\,,
\]
\be
q^3\approx {N_{HI}\Gamma_{12}\over N_0}
{b^2-\langle b_{bg}\rangle^2\over v_0^2}
\left({b\over\langle b_{bg}\rangle}\right)^{3/2}(1+z)^{-75/14}\,.
\label{shock}
\ee

Formation of absorbers with $b\leq b_{thr}=\beta_{thr}\langle
b_{bg}\rangle$
is accompanied by adiabatic or weak shock compression of
baryonic component. Assuming that the compression of baryons is
described by the polytropic equation of state with
$\gamma_b=5/3$, we can expect that for recently formed absorbers
with $b\leq b_{thr}$, $F_s(z)\approx \langle F_{bg}(z)\rangle$,
\be
q\approx {N_{HI}\Gamma_{12}\over
N_0}\beta^{-3/2} (1+ z)^{-61/14},\quad \delta_b\approx
\beta^3\,,
\label{adiab}
\ee
and for long lived absorbers (\ref{dltth})
\be
q\approx {N_{HI}\Gamma_{12}\over N_0}\beta^{-2}
(1+z)^{-61/14},\quad \delta_b\approx \beta^{7/2}\,.
\label{expand}
\ee

The relations (\ref{shock}) -- (\ref{expand}) determine the
dimensionless column density of DM component corrected for the
impact of gaseous pressure. These relations can be successfully
applied to absorbers formed due to adiabatic and strong shock
compression with various degrees of relaxation. However, the
boundary between these limiting cases must be established {\it a
priory}. So, to discriminate absorbers described by
(\ref{shock}) and  (\ref{adiab}, \ref{expand}), we use the
threshold Doppler parameter, $b_{thr}$ which, in fact,
characterizes the Mach number of the inflowing matter. Thus,
absorbers with $b\leq b_{thr}$, \be b_{thr}=\beta_{thr}\langle
b_{bg}(z)\rangle,\quad \beta_{thr} \approx 1.5 - 2\,,
\label{thr} \ee can be conveniently considered as the
adiabatically compressed while absorbers with $b\geq b_{thr}$
are considered as formed by shock compression and they contain
strongly relaxed matter.

The precision of these estimates is moderate and the some
uncertainties are generated by the poorly known $\Gamma_\gamma$
and the parameters $\Theta_\Phi$ and $\Theta_H$, which vary --
randomly and systematically -- from absorber to absorber.
Estimates of these uncertainties can be obtained from the
analysis of the derived absorbers' characteristics.
However, the main uncertainties in the estimates of $q$ come
from the unknown $\cos\theta$, and, for strongly relaxed shock
compressed absorbers, from modulations of $b_{bg}$.

Indeed, for adiabatically compressed absorbers $q\propto cos
\theta$ and Eqs.(\ref{adiab}, \ref{expand}) underestimate $q$
for $\cos\theta\geq\langle\cos\theta\rangle=0.5$ and
overestimate it for $\cos\theta\leq\langle\cos\theta\rangle$. To
reveal and to correct the most serious uncertainties we will use
the condition $d_{abs}/d_v\leq 1$ where the real, $d_{abs}$, and
the redshift, $d_v$, size of absorbers were introduced in
(\ref{delr}, \ref{br}).  For absorbers with $d_{abs} \geq d_v$
we will substitute the 'true' column density, $q_t$ defined by
relation \be q_t=q~d_v/d_{abs},\quad d_{abs}\approx d_v,
\label{correct} \ee instead of the measured one (\ref{adiab}) or
(\ref{expand}). Perhaps, more detailed reconstruction of the
shape of absorption lines could allow us to reveal stronger
deviations from the Doppler profile and, so, to identify the
influence of absorbers orientation and macroscopic velocities.

For shock compressed absorbers the influence of $\cos\theta$ is
not so strong as $q\propto\cos^{1/3}\theta$ and the expected
modulation of $b_{bg}$ is more important. Indeed, these
absorbers are formed owing to the merging which is more probable
in slowly expanded regions with $b_{bg}\geq \langle
b_{bg}\rangle$. Thus, for absorbers with $b\geq b_{thr}$,
$d_{abs}\geq d_v$, we will also correct $q\, \&\, d_{abs}$ by
relation (\ref{correct}).

These corrections essentially improve estimates of the
correlation function of initial velocity field.

\subsection{Regular and random variations of absorbers'
characteristics}

The most fundamental characteristic of absorbers is their DM column
density, $\zeta(q,z)\propto q(1+z)^2$ (\ref{xit}).  It depends on
the process of formation and merging of pancakes, is only weakly
sensitive to the action of random factors and defines the regular
redshift variations of absorbers characteristics.

For shock compressed absorbers, the evolutionary history of
each pancake and the action of random factors discussed in
the previous subsections are integrated in the entropy of
the baryonic component,
\be
S_b=\ln F_s(z)=S_{bg}+2/3\ln(\beta^3/\delta_b)\,.
\label{sbar}
\ee
If the structure of a relaxed DM pancake can be described by
the polytropic equation of state with the effective power index
$\gamma_m$ then we can introduce also the entropy of DM
component, $S_m$. For probable value $\gamma_m \approx 1.5 - 2$,
entropies of DM and baryonic components are quite similar to
each other, $S_m\approx S_b$. This means that for the strongly
relaxed shock compressed absorbers the evolutionary history is
characterized quite well by two functions, $\zeta$ and $S_b$.

For adiabatically compressed absorbers the baryonic entropy is
identical to the background one given by (\ref{sbg}) while the
observed $b$ and $N_{HI}$ depend upon the distribution of the
compressed DM component. For such absorbers the PDFs and the
random scatter of observed characteristics are defined mainly by
variations of the expansion rate and the background density and
temperature. These characteristics as well as the entropy and
overdensity of compressed DM component, $S_m\,\&\, \delta_m$,
now cannot be derived from observational data with a reasonable
reliability. This problem deserves further investigations.

\subsection{Reconstruction of the initial power spectrum}

The basic relation of Zel'dovich theory of gravitational
instability can be suitably written for the difference of
coordinates of two particles, \be \Delta r_i={l_v\tau(z)\over
1+z}[q_i/\tau(z)-\Delta S_i(q)/l_v], \quad i=1, 2, 3\,,
\label{zsld} \ee where $r_i$ are the Euler coordinates of
particles, $l_v$, $\tau(z)$ and $q_i$ were introduced in
(\ref{lv}), (\ref{xiv0}) \& (\ref{Btau}) and $S_i(q)$ is a
random displacement of a particle with respect to its
unperturbed position. As is seen from this relation, for
pancakes $\Delta S\approx l_vq/\tau(z)$ and, so, some
statistical characteristics of $\Delta S$ can be obtained
by measuring $q/\tau(z)$.

Using the measured redshift, $z$, and DM column density of
absorbers, $q$, we determine the cumulative PDF of absorbers
$W_{obs}[>q/\tau(z)]$ and, for each $q/\tau(z)$, we compute
$\langle q\rangle$ and $\sigma_q^2=\langle q^2\rangle-\langle
q\rangle^2$. For a chosen $W_q(\zeta)$ (\ref{qpdf}), we solve
numerically the equation
\be
W_{obs}[<q/\tau(z)]=W_q(\zeta)={\rm erf}^2(\zeta)\,,
\label{eq1}
\ee
with respect to $\zeta(q,\tau)$ and, thus, we obtain the
function
\be
1-\xi_v(q)={q^2\over 4\tau^2\zeta(q)}\,.
\label{eq11}
\ee
For the most interesting range $q/\tau(z)\ll1, \zeta\ll 1$
we have
\be
W_q(\zeta)\approx 4\zeta^2/\pi,\quad
1-\xi_v\propto q^2/\sqrt{W_q(\zeta)}\,.
\label{wqz}
\ee

The same approach can be applied to the absorbers separation,
$d_{sep}$, and $\zeta_{rd}=\zeta(q_{rd},z),\,\&\,q_{rd}$
introduced by (\ref{sep_mns})\,. In this case instead of
(\ref{eq1}) and (\ref{wqz}) we have
\[
W_{obs}[<d_{sep}/l_v\tau(z)]=W_{sep}(<\zeta_{rd})\approx
\]
\be
1-1.778\exp(-0.634\zeta_{rd})[1-0.425\exp(-0.634\zeta_{rd})]\,,
\label{eq21}
\ee
\[
1-\xi_v(q_{rd})={q_{rd}^2\over 4\tau^2\zeta(q_{rd})}\,,
\]
and, for $q_{rd}/\tau(z)\ll 1, \zeta_{rd}\ll 1$, we get
\[
W_{sep}(\zeta_{rd})\approx 0.17\zeta_{rd},\quad 1-\xi_v\propto
q^2_{rd}/W_{sep}\,.
\]

The observed functions $P_{obs}(q)$ and $P_{obs}(d_{sep})$
can be compared with corresponding expectations (\ref{qpdf})
and (\ref{sep_mns}). The correlation function $\xi_v(q)$
derived from (\ref{eq11}) and (\ref{eq21}) can be compared
with the reference function $\xi_{CDM}(q)$ (\ref{xiv}).

\section{Model dependent statistical characteristics of
absorbers}

In this Section, evolution of the basic model dependent
characteristics of absorbers is discussed in the $\Lambda$CDM
cosmological model (\ref{basic}), and for the background
temperature, $T_{bg}$, the Doppler parameter, $b_{bg}$, and the
entropy function $F_{bg}$ given by (\ref{bg}) and (\ref{sbg})\,.

\subsection{Parameters of the model}

The model of absorbers discussed in Sec. 3 includes poorly known
random parameters, $\Theta_{bg},\,\Theta_\Phi\,\&\, \Theta_x$, which
cannot be estimated {\it a priory}, this leads to a moderate random
scatter of derived absorbers characteristics. Further on we will
assume that
\be
\Theta_{bg}=\Theta_\Phi=\Theta_x=1\,.
\label{Theta}
\ee
However, as was discussed in Sec. 3.7, the main sources of
uncertainty  in (\ref{shock} -- \ref{expand}) are the random
orientation of absorbers with respect to the line of sight, measured
by $\cos~\theta$ and, for strongly relaxed shock compressed
absorbers, random spatial modulation of the expansion rate and the
background temperature, $b_{bg}\geq \langle b_{bg} \rangle$. Large
distortions of the derived parameters of absorbers can be revealed
and partly corrected using the relation (\ref{correct}). However
moderate distortions of the same parameters that can not be easily
corrected restrict the precision of our approach.

The poorly known radiative ionization rate, $\Gamma_{12}$,
(\ref{ioniz})\, is also an important source of uncertainty. As was
noted in Sec. 2.1, now there are approximate estimates of the UV
background produced by the observed QSOs but the expected ionization
rate should be corrected for the absorption and reemission of UV
radiation by the gas compressed within high density clouds (Haardt
\& Madau 1996) and for the additional emission of UV radiations by
galaxies (at $z\leq 2$) and at $z\geq 3$ by poorly  known sources
such as Ly--$\alpha$ emitters (Boksenberg, Sargent\,\&\,Rauch 2003;
Giavalisco et al. 2004; Ouchi 2005).

Below we will describe the ionization rate by the expression \be
\Gamma_{12}(z)\approx G_0\left({1+z\over 4}\right)^
{p_\gamma}\exp\left[-{(z-z_\gamma)^2\over 2\sigma^2_\gamma}
\right]\,, \label{g12} \ee
\[
G_0=4.3,\quad z_\gamma=1.0,\quad \sigma_\gamma=1.58,\quad
p_\gamma=1.5\,,
\]
where the choice of $p_\gamma, z_\gamma$ and $\sigma_\gamma$
corrects $\Gamma_{12}(z)$ for the impact of additional sources
of radiation.

Relatively small value of $\Gamma_{12}(2)\approx 3$ (\ref{g12})
is close to the observational estimates of Scott et al. (2002)
and $\Gamma_{12}(5)\approx 0.2$ is similar to the estimates of
Fan et al. (2002, 2004). It shows that at $z\sim 2$ the UV
background was probably overestimated in Haardt \& Madau (1996)
and Demia\'nski \& Doroshkevich (2004b)\,. Indeed, relatively
small observed $\langle N_{HI}\rangle \approx 10^{13.4} cm^{-2}$,
with $\lg N_{HI}\leq 15$, and moderate number of lines
$N_{line}=226$ with $lg N_{HI}\geq 15$ in our sample, shows
that the absorption of the UV background by $HI$ cannot be
very important. However, stronger absorption by $HeII$ found
by Levshakov et al. (2003) can distort the spectrum of the
UV background and decrease $G_0$ and the background
temperature, $T_{bg}\,\&\, b_{bg}$ (\ref{bg}). Below the fit
(\ref{g12}) will be tested by comparing the theoretically
expected and derived functions, $\xi_v(q)$ and $\xi_v(q_s),
\,\langle\zeta(z)\rangle \approx 0.82$ and $\langle l_v q(z)
\rangle\approx 0.44 \langle d_{sep}(z)\rangle$.

For such $\Gamma_{12}(z)$ we get for the Gunn--Peterson optical
depth, $\tau_{GP}(z)$,
\be
\tau_{GP}(2)\approx 0.045f^2_{hom},\quad \tau_{GP}(5)
\approx 5f^2_{hom}\,,
\label{gp2}
\ee
where $f_{hom}$ is the
fraction of homogeneously distributed matter. Estimates obtained
in Sec. 4.3 show that at $z\leq 4, ~~f_{hom}\approx 0.5$.
However, at such redshifts $\tau_{GP}$ is measured for regions
with suppressed lines where larger $\Gamma_{12}$ can be
expected. For large $z\geq 4$ we can take $f_{hom}\sim 1$ with
large scatter.

The choice of $\Gamma_{12}(z)$ (\ref{g12}) coincides with
\[
\tau_0\approx 0.19,\quad \sigma_8\approx 0.9\,,
\]
what agrees quite
well with $\sigma_8$ derived by Spergel et al. (2003, 2006) and with
the independent estimate (\ref{mnsbhs}). The final results only
weakly depend upon the threshold parameter $b_{thr}\approx (1.5-
2)b_{bg}$ discriminating between adiabatically and shock
compressed absorbers. Here we use $b_{thr}\approx 1.5b_{bg}$\,.

Such choice of the model parameters allow us to obtain
reasonable description of properties of absorbers for the
selected sample. Variations of limits used for the sample
selection lead to moderate variations of parameters (\ref{g12}).

\subsection{Correlation functions of the initial velocity
field}

\subsubsection{Correlation functions derived from absorber
separations}

Using the method described in Sec. 3.9  and characteristics of
separations between absorbers obtained in Sec. 2.2 we can
estimate also the correlation function of the initial velocity
field, $\xi_v(q_{rd})$\,. This approach uses only the measured
redshifts and, so, the derived function does not depend upon the
measured $b\,\&\,N_{HI}$ and the model of absorbers discussed in
Sec. 2. However, the theoretically derived PDFs (\ref{sep_mns})
and (\ref{eq21}) become unreliable at small separations and, as
was noted in Sec. 2.2, already at redshifts $z\sim$ 3 the
blending of absorbers becomes essential what distorts the
derived function $\xi_v(q_{rd})$.

\begin{figure}
\centering 
\epsfxsize=7.cm
\epsfbox{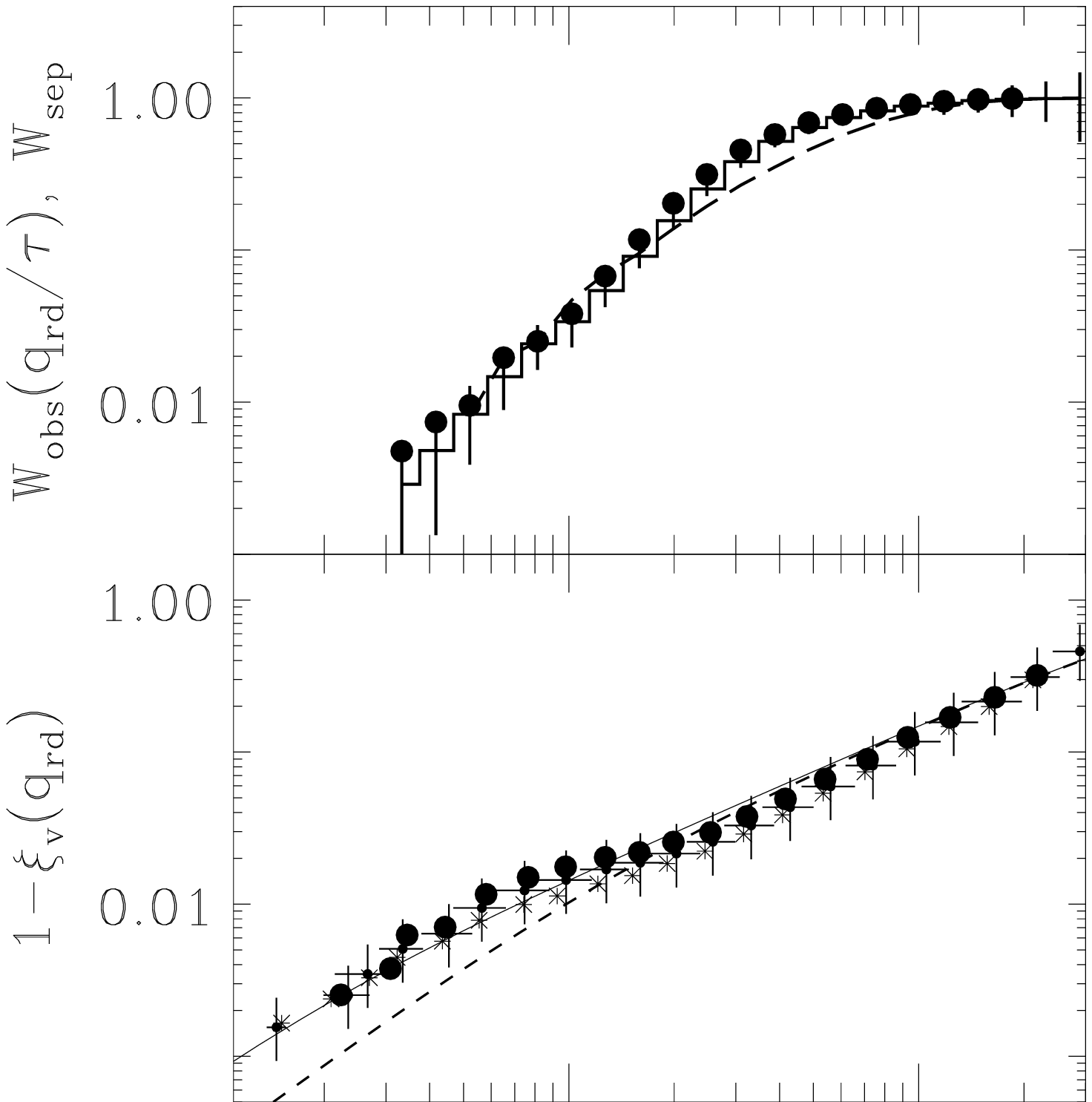}
\vspace{1.cm}
\caption{Top panel: the cumulative PDF, $W_{obs}(q_{rd}/\tau)$,
together with the fit $W_{sep}(\zeta_{rd})$ (\ref{eq21}) (dashed
line).
Bottom panel: the correlation function of initial velocity field,
$1-\xi_v(q_{rd})$, together with the theoretical fits (\ref{xiv})
(dashed line) and the fit (\ref{sep_fit}) (solid line).
Points and stars show the same functions found for the sample
of 14 QSOs and for the sample of 7\,411 separations in 19
spectra, respectively.
}
\label{sep100}
\end{figure}

The observed cumulative PDF, in redshift space,
$W_{obs}[q_{rd}/\tau(z)]$, and the reconstructed correlation
function $1-\xi_v(q_{rd})$ (\ref{eq21}) are plotted in Fig.
\ref{sep100} for the samples of 19 QSOs (6\,251 and 7\,411
separations) and 14 QSOs (3\,660 separations). For these samples we
have, respectively,
\[
W_{obs}[<q_{rd}/\tau(z)]/W_{sep}(<\zeta_{fit})\approx 1.07\pm
0.19,\quad 1.03\pm 0.2\,,
\]
\be {1-\xi_v(q_{rd})\over 1-\xi_{fit}(q_{rd})}\approx 0.95\pm
0.25,\quad 1\pm 0.2\,,
\label{sep_fit}
\ee
\[
1-\xi_{fit}(q_{rd})={1.5q_{rd}^2\over\sqrt{2.25q_{rd}^4+q_{rd}^2+
p_s^{1.4}q_{rd}^{0.6}}},\quad p_s=2.1\cdot 10^{-3},
\]
\[
0.5\geq q\geq 1.4\cdot 10^{-3},\quad 17h^{-1}{\rm Mpc}\geq
l_vq_{rd}\geq 0.03h^{-1}{\rm Mpc} \,.
\]
where $\zeta_{fit}=\zeta(\xi_{fit},\tau)$ is given by (\ref{xit}).

As is seen from  Fig. \ref{sep100}, at $q_{rd}\geq  10^{-2}$,
$l_vq_{rd}\geq 0.33h^{-1}$Mpc the derived correlation function
coincides with the standard CDM -- like one (\ref{xiv})  for heavy
DM particles, with $q_0\leq 10^{-3}, M_{DM}\geq$ 1MeV\,. Similar
results can be found also with the PDF (\ref{Lsep}) what indicates
the weak sensitivity of the derived correlation function
$\zeta_{fit}$ on the detailed shape of the used PDF.

At small scales, $10^{-3}\leq q_{rd}\leq  10^{-2}$, we see an
excess of power with respect to the reference function $\xi_v$
(\ref{xiv}). This excess is caused by the deficit of observed
absorbers with small separation which increases the derived
function, $1-\xi_v(q_{rd})\propto W_{sep}^{-1}$ (see Sec. 3.9).
In this range of $q_{rd}$ the number of measured separations is
limited, $N_{sep}\approx 320$ and $N_{sep}\approx 170$ for the
samples of 19 and 14 QSOs, respectively, what decreases
reliability of this result. To test impact of this factor we
calculated the function $\xi_v(q_{rd})$ for the sample of 7\,411
separations obtained without placing any restrictions on the
properties of absorbers. For this extended sample the number of
small separations increases up to 630 but the difference between
the derived and reference functions, $\xi_v(q_{rd})$ and
$\xi_{CDM}(q_{rd})$, at $q_{rd}\leq 10^{-2}$ remains the same.

These results indicate that possibilities and applicability
of this approach are limited. Indeed, the positions of absorbers
are measured in the redshift space with a typical error $\Delta
z \sim 5 \cdot 10^{-5}$, $\Delta q_{rd}\sim 3\cdot 10^{-2} [4/(1+z)
]^{3/2}$ comparable with the size of absorbers as measured by
their Doppler parameter (\ref{mnsbhs}), $(1+z) \langle d_v
\rangle/l_v\sim 3\cdot 10^{-2}[4/(1+z)]^{1/2}$\,. These factors
lead to the artificial blending of close absorbers, distort
their characteristics and decrease reliability of the estimates
(\ref{sep_fit}) for $q_{rd}\leq 10^{-2}$.

\subsubsection{Correlation functions derived from DM column
density of absorbers}

The same correlation function of the initial velocity field,
$\xi_v(q)$, can be also found from estimates of the DM column
density of absorbers, $q$.  Here we use a more complex procedure
of determination of $q$ from the observed $z,\,b,\,\&\,N_{HI}$
and poorly known $\Gamma_{12}$ (\ref{g12}) what decreases its
reliability. On the other hand, comparison of the functions
$\xi_v$ found with two different approaches allows us to test
the model of absorbers discussed in Secs. 3 and 4.1\,.

\begin{figure}
\centering \epsfxsize=7.cm
\epsfbox{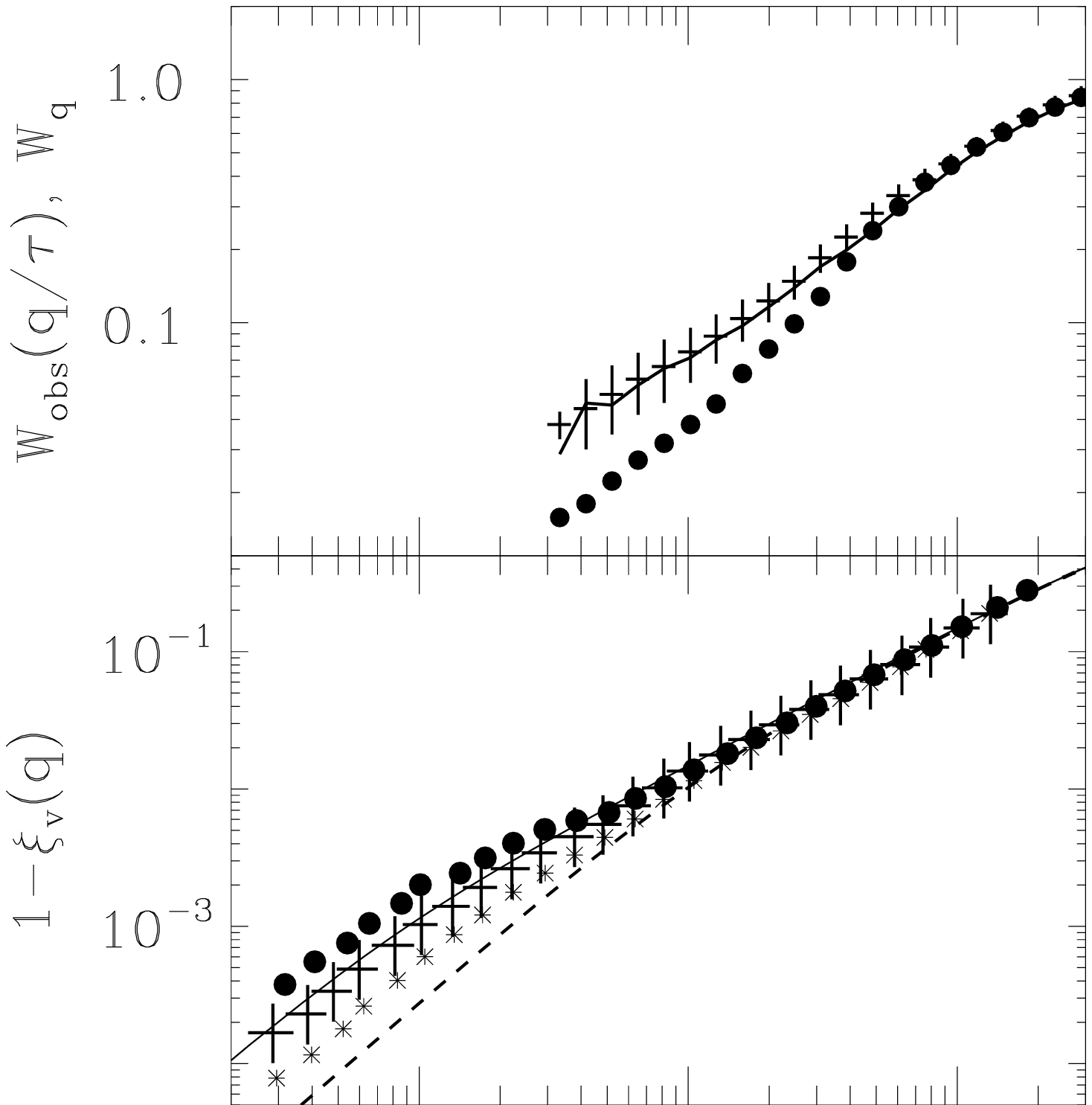}
\vspace{1.cm}
\caption{Top panel: the cumulative PDF, $W_{obs}(q/\tau)$,
together with fit (\ref{eq1}) (long dashed line). Bottom
panel: the correlation function of initial velocity field,
$1-\xi_v(q)$, together with the theoretical fit (\ref{xiv})
(dashed line) and the fit (\ref{cor_fit}) (solid line)\,.
Points and stars show the same functions found for the sample
of 14 QSOs and for the sample of 7\,430 separations.
}
\label{cor100}
\end{figure}

The observed cumulative PDF, $W_{obs}(q/\tau)$, and the
reconstructed correlation function $1-\xi_v(q)$ are plotted in
Fig. \ref{cor100} for the samples of 19 QSOs with 6\,270 and
7\,430 absorbers and 14 QSOs (3\,674 absorbers). For these
samples we have, respectively,
\[
W_{obs}(<q/\tau)/W_q(\zeta_{fit})\approx 1.\pm 0.07,\quad
0.9\pm 0.2\,,
\]
\be
{1-\xi_v(q)\over 1-\xi_{fit}(q)}\approx 0.95\pm 0.15,\quad
1.1\pm 0.2\,,
\label{cor_fit}
\ee
\[
1-\xi_{fit}(q)\approx {1.5q^2\over\sqrt{2.25q^4+q^2+p_q^{1.4}
q^{0.6}}},\quad p_q=0.8\cdot 10^{-3}\,,
\]
\[
10^{-4}\leq q\leq 0.3,\quad 3h^{-1}{\rm kpc} \leq l_vq\leq 9.4
h^{-1}{\rm Mpc}\,,
\]
where $\zeta_{fit}=\zeta(\xi_{fit},\tau)$ is given by (\ref{xit}).

As is seen from Fig. 4, at $q\leq 5\cdot 10^{-3}$ the derived
correlation function (\ref{cor_fit}) differs from the standard
CDM -- like one (\ref{xiv})\,. For these $q$ both samples of
absorbers are quite representative with $N_{abs}\approx 1590$
and $N_{abs}\approx 850$ for samples of 19 and 14 QSOs,
respectively. However, as was noted above, a more complex
procedure of determination of $\xi_v$ decreases its reliability.

At these scales  the difference between the derived and
reference correlation functions, $\xi_v(q)$ and $\xi_{CDM}$
(\ref{xiv}), could be mainly caused by the deficit of weaker
absorbers (Sec. 3.9) because $1-\xi_v(q) \propto
W_q^{-1/2}$. As before, to test the impact of this factor we
calculated the function $\xi_v(q)$ for the sample of 7\,430
absorbers with $\lg N_{HI}\leq 15$. For such sample the
difference between the derived and reference functions,
$\xi_v(q)$ and $\xi_{CDM}(q)$, at $q\leq 10^{-3}$ decreases and
the correlation function is fitted by the expression
\be
1-\xi_{fit}(q)={1.5q^2\over\sqrt{2.25q^4+q^2+p_f^{1.4}
q_{rd}^{0.6}}},\quad p_f=4\cdot 10^{-3}\,.
\label{cor_cor}
\ee
The functional forms of expressions (\ref{xiv}), (\ref{cor_fit})
and (\ref{cor_cor}) are identical and they differ only by the
values of fit parameters $p_q\approx 0.07p_0,\, p_f\approx 0.3
p_0$. This means that the difference between $\xi_v$
(\ref{cor_fit}) and $\xi_{CDM}$ (\ref{xiv}) could be mainly
related to a possible incompleteness of the observed samples
of weaker absorbers, to the limited precision of measurements
of $b\,\&\,N_{HI}$ and to the limited  precision of our model
in describing such absorbers (see discussion in Sec. 3.7).

The correlation function (\ref{cor_cor}) is quite similar to the
reference one and their difference is in the range of
observational errors. This fact demonstrates that probably the
CDM like initial power spectrum can be traced at least down to
$q\sim 10^{-4}$, $l_vq\sim 3h^{-1}$ kpc. This means also that
$q_0\leq 10^{-4}$, $m_0\geq 25$ and the effective mass of the DM
particles $M_{DM}\geq 100$ MeV. However, the function
(\ref{cor_cor}) is derived from inhomogeneous samples and
therefore its reliability is in question.

At larger scales, $q\geq 5\cdot 10^{-3},\,l_vq\geq 0.15 h^{-1}$
Mpc, results obtained from the analysis of both characteristics
of absorbers are quite similar: \be {1-\xi_v(q_{rd})\over
1-\xi_{CDM}(q_{rd})}\approx 1.08\pm 0.25, \quad 1.06\pm 0.17\,,
\label{sep_lg} \ee \be {1-\xi_v(q)\over 1-\xi_{CDM}(q)}\approx
1.0\pm 0.2,\quad 1.09\pm 0.16\,. \label{cor_lg} \ee These
results confirm the CDM -- like type of the initial power
spectrum down to scales $l_vq\geq 0.15 h^{-1}$Mpc what extends
conclusions of Croft et al. (2002), Viel et al. (2004b);
McDonald et al. (2004) and Zaroubi et al. 2005. It also
demonstrates the self consistency of the adopted model
of absorbers.

It is important, that similar results are found for both samples
of 19 QSOs with 6\,270 absorbers and of 14 QSOs with 3\,674
absorbers. The samples used are compiled from spectra observed
with different instruments and resolutions and the parameters of
absorbers were found with different codes what increases their
possible non homogeneity. Under these conditions, the stability
of our results demonstrates their objectivity.

Reconstruction of the initial power spectrum with the help
of the relations (\ref{xivv}) shows that in the range of errors
the measured and CDM--like power spectra are quite similar each
other. At larger $k$, $k/k_0\geq 100$, there is some excess of
the power but estimates become unstable because
of the limited range of measured $q$. Investigations of this
important  problem should be continued with more homogeneous
sample of observed spectra.

\subsection{Statistical characteristics of the full sample
of absorbers}

\begin{figure}
\centering 
\epsfxsize=7.cm
\epsfbox{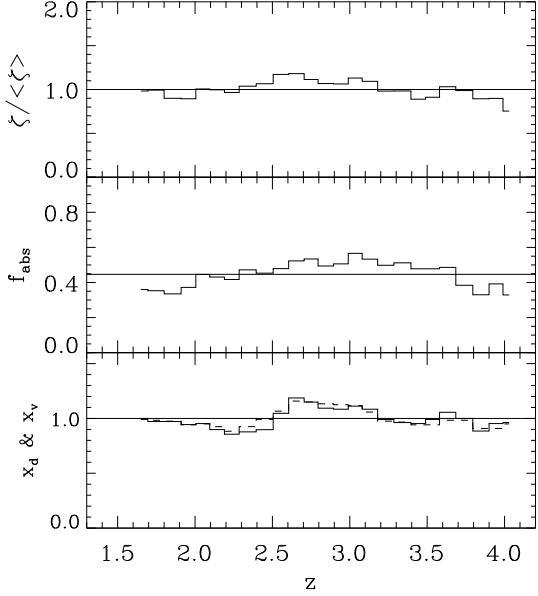}
\vspace{1.cm}
\caption{Functions $\zeta(z)/\langle\zeta\rangle$,
(top panel), the matter fraction accumulated by absorbers,
$f_{abs}(z)$, (middle panels), and the real and redshift sizes
of absorbers, $x_d$ (\ref{Pdabs}), and $x_v=d_v(z)z_4^{3/2}/
\langle d_vz_4^{3/2}\rangle$, (bottom panel, solid and dashed
lines).
}
\label{mns_ful}
\end{figure}

\begin{figure}
\centering 
\epsfxsize=7.cm
\epsfbox{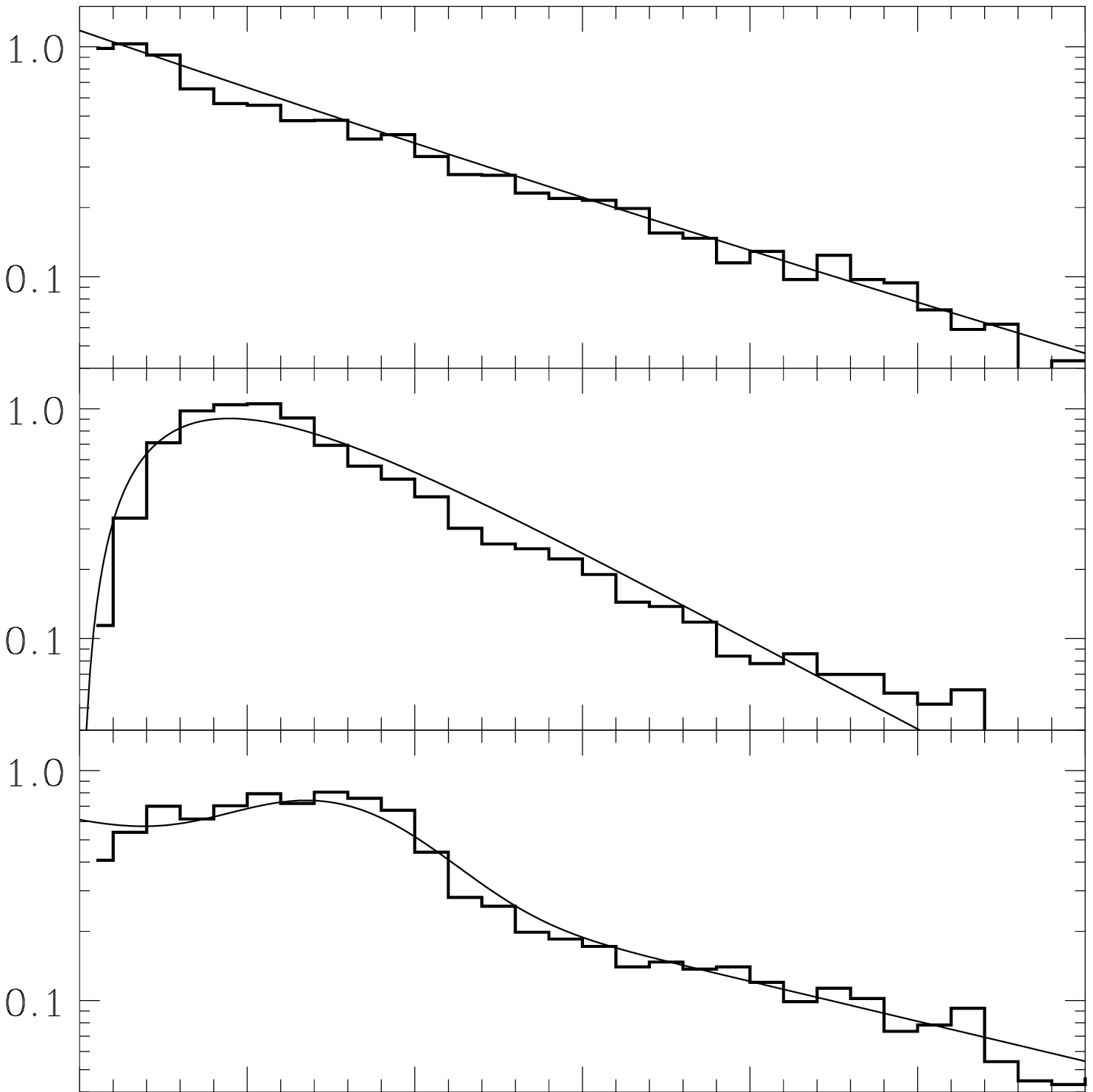}
\vspace{1.cm}
\caption{PDF for the DM column density, $P(x_\zeta)$, (top panel),
the absorber separation, $P(x_{rd})$, (middle panel),
and for the absorber thickness, $P(x_d)$ (bottom panel) are
plotted together with fits (\ref{zfit}), (\ref{spfit}) and
(\ref{Pdabs}). }
\label{hstfl}
\end{figure}

For the sample of 6\,270 absorbers the redshift variations of
the mean DM column density, $\langle\zeta\rangle$, the mean
fraction of DM component accumulated by absorbers, $\langle
f_{abs}\rangle$, and the real and redshift sizes of absorbers
along the line of sight, $\langle d_{abs}z_4^{3/2} \rangle$, and
$\langle d_vz_4^{3/2}\rangle$, are plotted in Fig. \ref{mns_ful}.
The PDFs for the functions $\zeta$ and $d_{abs}z_4^{3/2}$ are
plotted in Fig. \ref{hstfl}. For $\sim$ 1\,500 absorbers of this
sample the parameter $q$ was corrected as described in Sec. 3.7
(Eq. (\ref{correct})). As was expected, the fraction of such
absorbers ($\sim$ 25\%) is close to the probability
$0\leq\cos\theta\leq 0.25$ to find absorbers oriented along the
line of sight.

The mean DM column density of absorbers, $\langle q\rangle$ or
$\langle\zeta\rangle$, is the most stable characteristic of the
sample. In principle, $\langle\xi\rangle$ does not change due to
the formation and merging of absorbers and due to their
transverse compression and/or expansion (DD04). It depends upon
the ionization rate, $\Gamma_{12}$, the amplitude of initial
perturbations, $\tau_0$ or $\sigma_8$, and upon the shape of the
correlation function of initial perturbations, $\xi_v$, or the
initial power spectrum, $p(k)$. It also weakly depends upon the
parameter $b_{thr}$ used to discriminate absorbers formed by the
adiabatic and shock compression. The differences between the
expected and measured $\langle\xi\rangle$ characterize, in fact,
the completeness and representativity of the samples, the
scatter of the function $\Gamma_{12}$ and the influence of
disregarded factors such as
$\Theta_{bg},\,\Theta_\Phi\,\&\,\Theta_x$.

For the sample under consideration the redshift variations of
the measured $\langle\xi(z)\rangle$ around the mean values are
moderate,
\be
\langle\zeta\rangle \approx 0.82\pm 0.08,\quad
\langle qz_4^2\rangle \approx (1.8\pm 0.2)\cdot 10^{-2}\,.
\label{qttt}
\ee
At small and larger redshifts, $z\leq 2$ and
$z\geq 3.7$, limited statistic of absorbers decreases
reliability of our estimates. The measured $\langle\zeta(z)
\rangle$ is close to the theoretically expected value (\ref{qpdf})
what verifies the choice of the ionizing rate $\Gamma_{12}(z)$ in
(\ref{g12}) and the amplitude of initial perturbations,
$\tau_0=0.19, \sigma_8 \approx 0.9$\,.

The PDF of the DM column density, $P(\zeta)$ plotted in Fig.
\ref{hstfl} is fitted with a scatter $\sim$ 10\% by the function
\be
P(x_\zeta)=1.2\exp(-x_\zeta){\rm erf}(\sqrt{x_\zeta})/
\sqrt{x_\zeta},\quad x_\zeta=\zeta(q,z)/\langle\zeta\rangle\,,
\label{zfit}
\ee
which is very close to the theoretical relation
(\ref{qpdf})\,. This result verifies the self consistency of the
physical model used here and the assumed Gaussianity of initial
perturbations.

As is well known, for the full sample the HI column density and
Doppler parameter are weakly correlated, and for our sample
their linear correlation coefficient is \be R_{b HI}=[\langle
bN_{HI}^*\rangle-\langle b\rangle\langle
N_{HI}^*\rangle]/\sigma_b\sigma_{HI}^*\approx 0.16\,,
\label{eq111} \ee where $N_{HI}^*=N_{HI}/z_4^2$. At the same
time, the DM column density, $\zeta$, is correlated with both
the HI column density and Doppler parameter, and the linear
correlation coefficients defined in the same manner as
(\ref{eq111}) are 
\be 
R_{\zeta_b}\approx 0.34,\quad R_{\zeta_HI}\approx 0.72\,. 
\label{corcoef} 
\ee

The PDF of observed absorber separations, $P(x_{rd})$, plotted
in Fig. \ref{hstfl} is fitted with a scatter $\sim$ 13\% by the
function \be
P(x_{rd})=3.56\exp(-1.6x_{rd}){\rm erf}^4
(\sqrt{1.6x_{rd}})/\sqrt{x_{rd}}\,, \label{spfit} \ee where
$x_{rd}=\zeta_{rd}(d_{sep},z)/\langle\zeta_{rd}\rangle$,
$\langle\zeta_{rd}\rangle\approx 1.5$ and $\zeta_{rd}(d_{sep}, z)$
was introduced in (\ref{sep_mns}). At small and large $x_{rd}$ this
fit differs from the theoretically expected one for redshift space
(\ref{sep_mns}) but it is quite similar to the fit (\ref{Lsep}) for
the PDF of separations in Lagrangian space. Perhaps, this fact can
be related to the unexpectedly moderate influence of the peculiar
velocities of absorbers. However, it can be partly caused by
peculiarities of our samples.

Our model of absorbers (see Sec. 3) allows one to estimate
roughly the real size of baryonic distribution across the
absorber, $d_{abs}$, (\ref{delr})\,.  For the full sample we
have \be \langle d_{abs}^*\rangle=\langle
d_{abs}z_4^{3/2}\rangle \approx (0.11\pm 0.01)h^{-1}{\rm Mpc}\,,
\label{dabs} \ee
\[
\langle d_v^*\rangle=\langle d_vz_4^{3/2}\rangle\approx
(0.12\pm 0.01)h^{-1}{\rm Mpc}\,,
\]
and both sizes increase with time. The PDF of the real size of
baryonic pancakes is roughly fitted by:
\be P(x_d)\approx 0.6\exp(-0.8x_d)+0.4\exp[-(0.73-x_d)^2/0.15]\,,
\label{Pdabs}
\ee
\[
x_d=d_{abs}^*/\langle d_{abs}^*\rangle=d_{abs}z_4^{3/2}/\langle
d_{abs}z_4^{3/2}\rangle\,,
\]
with a scatter $\sim 12\%$.

For the expected mean transverse size of absorbers
(\ref{trsize}) and for model parameters (\ref{g12}) we have
for the mean proper size
\[
\langle l_vq_{tr}\rangle(1+z)^{-1}\approx 0.45 z_4^{-2}
(1+z)^{-1}h^{-1}{\rm Mpc}\,,
\]
what is consistent with recent direct estimates by Becker,
Sargent \& Raugh (2004) at $z\sim 3 - 3.5$
\[
l_vq_{tr}\approx (0.15 - 0.2)h^{-1}{\rm Mpc}\,.
\]
The exponential PDF of the transverse sizes (\ref{trsize}) and
its strong redshift dependence explain large scatter of the
sizes ($l_vq_{tr}\leq 1 h^{-1}$ Mpc) measured in many
observations of pairs of QSOs (see, e.g., discussion in Becker,
Sargent \& Raugh 2004).

The mean measured fraction of matter accumulated by absorbers is
\be
\langle f_{abs}(z)\rangle\approx 0.44\pm 0.07\,.
\label{fabs}
\ee
In spite of the limited applicability of the
one dimensional approach (\ref{fract}) and the limited precision
of our model of absorbers the measured fraction (\ref{fabs}) is
close to the theoretical expectation of the Zel'dovich theory
(\ref{ffr}) what verifies the choice of the model
characteristics (Sec. 4.1). The weak redshift variations of this
function agrees well with the self similar evolution of
absorbers.

\begin{table}
\caption{Mean parameters of adiabatically and shock
compressed absorbers}
\label{tbl2}
\begin{tabular}{lrr} 
\hline
   &adiabatic&shock\cr
\hline
$\langle lg N_{HI}/z_4^2\rangle$&$13.2\pm 0.07$&$13.5\pm 0.1$\cr
$\langle b\rangle km/s$ &$21\pm 0.9$&$48\pm 5$\cr
$\langle f_{n}\rangle$&$0.8\pm 0.06$&$0.2\pm0.05$\cr $\langle
f_{abs}\rangle$&$0.33\pm 0.05$&$0.11\pm 0.08$\cr
$\langle\zeta\rangle$&$0.77\pm 0.07$&$1.1\pm 0.2$\cr
$R_{\zeta_b}$&$0.39$&$0.42$\cr $R_{\zeta_HI}$&$0.75$&$0.77$\cr
$\langle\delta_b\rangle$&$\beta^{7/2}$& $(2\pm
0.3)z_4^{-2.5}$\cr $\langle S_b+2\ln z_4\rangle$&      &$2.3\pm
0.2$\cr $\langle d_{abs}z_4^{3/2}\rangle h^{-1}{\rm
Mpc}$&$0.09\pm 0.01$
 &$0.18\pm 0.05$\cr
$\langle d_vz_4^{3/2}\rangle h^{-1}{\rm Mpc}$&$0.1\pm 0.05
$&$0.22\pm 0.02$\cr
\hline
\end{tabular}

$z_4=(1+z)/4$, $f_n$ and $f_{abs}$ are the fraction of absorbers
and of matter in absorbers, $R_{\zeta_b}\,\&\,R_{\zeta_HI}$ are
the linear correlation coefficients of hydrogen and DM column
densities and Doppler parameter defined in the same manner as
(\ref{eq111}), $\delta_b$ and $S_b$ are the overdensity above
the mean density and relative entropy of compressed baryonic
component, $d_{abs}$ and $d_v$ are the real and redshift sizes
of absorbers along the line of sight.
\end{table}

\subsection{Adiabatically and shock compressed absorbers}

\begin{figure}
\centering
\epsfxsize=7.cm
\epsfbox{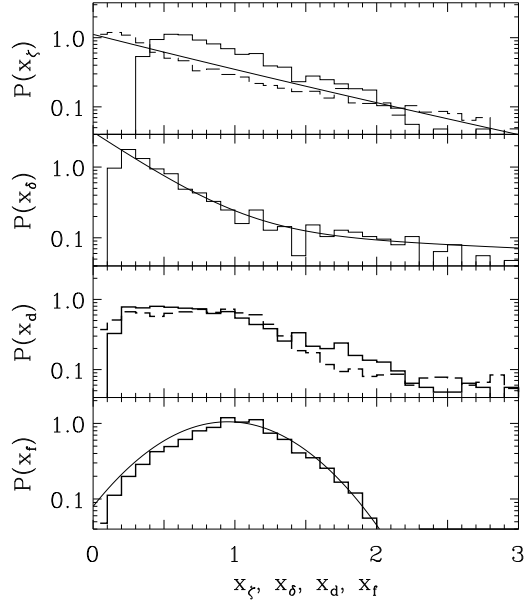}
\vspace{1.cm}
\caption{The PDFs for the DM column density, $P(x_\zeta)$,
overdensity, $P(x_\delta)$, and the real size across the
absorber, $P(x_d)$, (three top  panels) are plotted
for shock compressed (solid lines) and adiabatically
compressed (dashed lines) absorbers together with fits
(\ref{zfit}) and (\ref{shdelta}) . For shock compressed
absorbers, the PDFs for the baryonic entropy, $P(x_f)$,
is plotted in bottom panel with fit (\ref{entr}).
}
\label{hstful}
\end{figure}

To investigate the complicated evolution of absorbers in more
details, we compare subpopulations of adiabatically and
shock compressed absorbers. These subpopulations were separated
by comparison of the measured Doppler parameter, $b$, with the
background one $\langle b_{bg}\rangle$, (\ref{bg}). By
definition, absorbers with $b\geq b_{thr}=1.5 \langle b_{bg}
\rangle$ belong to  the subpopulation of shock compressed
and strongly relaxed absorbers, while absorbers with $b\leq
b_{thr}=1.5 \langle b_{bg}\rangle$ are considered as formed in a
course of adiabatic or weak shock compression. This discrimination
is not strict however and characteristics of subpopulations depend
upon the sample used in the analysis and the parameters of the
parameters of the background (\ref{bg},~\ref{g12}). This
classification allows one to
characterize reasonably well the observed absorbers and to
trace their evolutionary history. For both subpopulations,
redshift variations of the mean characteristics are listed
in Table 2 and some of the PDFs are plotted in Fig.\ref{hstful}.

For the sample under investigation $\sim$ 80\% of absorbers are
compressed adiabatically and they accumulate $\sim$ 80\% of the
compressed matter. These fractions, $f_n\,\&\,f_{abs}$, weakly
vary with redshift. The PDFs, $P(x_\zeta), x_\zeta=
\zeta/\langle\zeta\rangle$, plotted in Fig. \ref{hstful} for
both subpopulations are quite similar to each other and to the
PDFs (\ref{merg}) and $P(x_\zeta)$ (\ref{zfit}) plotted in Fig.
\ref{hstfl} for the full sample. It demonstrates that in wide
range of redshifts absorbers could be formed by both processes
while the cutoff at $\zeta=0.3\langle\zeta\rangle$ in the PDF
for shock compressed absorbers is imposed by the method used for
the discrimination of absorbers. For both samples, the
correlation coefficients $R_{\zeta_b}$ and $R_{\zeta_HI}$
defined in the same manner as (\ref{eq111}) are similar to each
other. These similarities verify the generic nature of absorbers
and indicate that they can be successfully combined into one
sample, what is consistent with expectations of the Zel'dovich
theory. For the subpopulation of shock compressed and strongly
relaxed absorbers $\langle\zeta\rangle\approx\langle\zeta_{mrg}
\rangle$ (\ref{merg}), as expected for merged absorbers, what
confirms the importance of merging in the process of absorbers'
evolution.

For both subpopulations $\langle b/b_{inf}\rangle\leq 1$, what
can be partly related to evaporation of high velocity particles
in the course of relaxation of the compressed matter. For both
subpopulations, the redshift size of absorbers along the line of
sight, $d_v$, is larger than the real size, $d_{abs}$ (Table 2).
The redshift dependence $d_{abs}\propto z^{-3/2}$ is consistent
with the evolution of their redshift size. The PDFs of the real
size of absorbers, $P(x_d)$, is plotted in Fig. \ref{hstful}.

For adiabatically compressed absorbers the entropy is the same
as for the background  while the overdensity,
$\delta_b=\beta^3$, is determined by  $b$ and $b_{bg}$. In
particular, for cold absorbers with $b\leq b_{bg}$ we have
$\delta_b\leq$ 1.

For subpopulation of strongly relaxed and shock compressed absorbers
both the entropy and the overdensity depend upon the complex
evolutionary history of absorbers and cannot be described by a
simple theoretical model. For such absorbers growth of the
overdensity with time can be explained by the action of several
factors such as successive merging and contribution of long lived
absorbers formed at redshifts $z_f$ larger than the observed one,
$z_{obs}\leq z_f$\,. For such absorbers the measured PDF
$P(x_\delta)$ is plotted in Fig. \ref{hstful} and it can be fitted
by a superposition of two exponential functions: \be
P(x_\delta)\approx 3\exp(-3.1x_\delta)+0.13\exp(-0.2x_\delta)\,,
\label{shdelta} \ee
\[
x_\delta=\delta_b (1+z)^{2.5}/\langle\delta_b (1+z)^{2.5}
\rangle\,.
\]
The entropy of strongly relaxed and shock compressed baryons
(\ref{sbar}) increases with time and its PDF plotted in Fig.
\ref{hstful} is well fitted by the Gauss function
\be
P(x_f)\approx \exp[-(x_f-0.95)^2/0.35]\,,
\label{entr}
\ee
\[
x_f=\ln[F_s(z)z_4^2]/\langle\ln[F_s(z)z_4^2]\rangle\,.
\]
Such PDF naturally arises when the entropy is generated by
the action of many random factors such as the shock waves
accompanying the successive merging of absorbers.

\subsection{Absorbers in rapidly and moderately expanded
regions}

As was discussed in Sec. 3.5, we expect that majority of absorbers
with small Doppler parameters, $b\leq b_{rap}\approx$ 23.5 km/s, and
especially with $b\leq \langle b_{bg}\rangle$ could be formed within
rapidly expanded regions and, so, they  can characterize some
properties of these regions. Some  characteristics of absorbers with
$b\leq b_{rap}$ are listed in Table 3 in comparison with the same
characteristics of adiabatically compressed absorbers with
$b_{thr}\geq b\geq b_{rap}$ situated within moderately expanded
regions.

The rapidly expanded regions accumulate $\langle f_n\rangle \sim$
50\% of adiabatically compressed absorbers and $\langle
f_{rap}\rangle\sim$ 30\% of adiabatically compressed matter. These
values practically do not depend on the redshift. The Doppler
parameter, $b$, and the DM column density, $\zeta_{rap}(q,z)$, also
weakly depend upon redshift. For this subpopulation we have  \be
\langle\zeta_{rap}\rangle\sim 0.5\langle\zeta\rangle,\quad \langle
f_{rap}\rangle\sim 0.25\langle f_{abs}\rangle\,. \label{zrap} \ee

As was noted in Sec. 2.2, the mean size of rapidly expanded regions,
$\langle D_{rap}\rangle$, as well as the mean separation of
absorbers within these regions, $\langle d_{sep}\rangle$, increase
with time in the same manner as the mean separation of absorbers for
the full sample (\ref{mnsbhs}). At $z_4\sim 1$ the typical mass
associated with rapidly expanded regions, $M_{rap}$
(\ref{tmass}), is in the range of galactic masses and it
increases with time $\propto (1+z)^{-3}$. This result is consistent
with the expected symmetry of positive and negative initial density
perturbations what leads to formation of both galaxies and rapidly
expanded regions.  These results are consistent with
theoretically expected ones (\ref{nrap}), what confirms the
interpretation of the complex shape of PDF $P_b(b)$ and
subpopulation of weak absorbers proposed in Sec. 3.5\,.

For the subpopulation of moderately expanded absorbers variations of
the Doppler parameter, $b$, are small, $R_{\zeta b}\ll 1$ and the DM
column density, $\zeta$, depends mainly upon the hydrogen column
density, $N_{HI}$. In contrast, for the rapidly expanded absorbers
the influence of both $b\, \&\,N_{HI}$ are equally important. In
spite of this difference, for both subpopulations the PDFs
$P(\zeta)$ are quite similar to each other and to the PDF
(\ref{zfit}) obtained for the full sample. This fact indicates that
the interaction of small and large scale perturbations changes
$\langle\zeta \rangle$ more strongly but only weakly influences the
shape of the PDFs $P(\zeta)$.

\begin{table}
\caption{Parameters of absorbers in rapidly  and moderately
expanded regions}
\label{tbl2}
\begin{tabular}{lrr} 
\hline
   &rapidly&moderately\cr
\hline $\langle lg N_{HI}/z_4^2\rangle$&$13.\pm 0.1$&$13.4\pm
0.1$\cr $\langle b\rangle km/s$ &$15.5\pm 0.6$&$27\pm 1.2$\cr
$\langle f_n\rangle$&$0.49\pm 0.05$&$0.51\pm 0.05$\cr $\langle
f_{abs}\rangle$&$(0.11\pm 0.05)$&$(0.22\pm 0.05)$\cr
$\langle\zeta\rangle$&$0.46\pm 0.04$&$1\pm 0.1$\cr
$R_{\zeta_b}$&$0.8$&$0.07$\cr $R_{\zeta_HI}$&$0.4$&$0.76$\cr
$\langle D_{reg}z_4^2\rangle h^{-1}{\rm Mpc}$&$(1.6\pm 0.2)
$&$(3.2\pm 0.4)$\cr $\langle d_{abs}z_4^{3/2}\rangle h^{-1}{\rm
Mpc}$&$(0.08\pm 0.01) $&$(0.1\pm 0.01)$\cr $\langle
d_vz_4^{3/2}\rangle h^{-1}{\rm Mpc}$&$(0.08\pm 0.003)
$&$(0.12\pm 0.005)$\cr \hline
\end{tabular}

$z_4=(1+z)/4$, $\langle f_n\rangle$ and $\langle f_{abs}
\rangle$ are the fraction  absorbers and of matter in absorbers,
$R_{\zeta_b}\,\&\,R_{\zeta_{HI/z_4^2}}$ are the linear
correlation coefficients of hydrogen and DM column densities and
Doppler parameter defined in the same manner as (\ref{eq111}),
$\langle D_{reg}\rangle$ is the mean size of rapidly and slowly
expanded regions, $\langle d_{sep}\rangle$ and $\langle
d_v\rangle$ are the mean absorber separation and their mean
redshift size along the line of sight,
\end{table}

For both subpopulations, the mean proper sizes of absorbers,
$\langle d_{abs}\rangle$, are similar (Table 3) but their PDFs
are quite different. Thus, for rapidly expanded regions the PDF
$P(x_d)$ is step--like and it is responsible for the bump at
$d_{abs}z_4^{3/2}\leq \langle d_{abs}z_4^{3/2} \rangle$ in the
PDF $P(x_d)$ plotted in Fig. \ref{hstfl}. This
distribution differs from the distribution of Doppler parameter,
$P(x_b)$, what suggests a complex internal structure of such
absorbers.

\subsection{Absorbers and properties of the background}

Some characteristics of absorbers could be used to estimate the
redshift variations of the mean properties of homogeneously
distributed hydrogen (see, e.g., Hui \& Gnedin 1997; Schaye et
al. 1999, 2000; McDonald et al. 2001). However, such estimates
are inevitably approximate and their significant scatter is
caused by the action of many random factors discussed above. To
obtain more stable results Schaye et al. (1999, 2000) consider a
cutoff at small $b$ in the distribution of $b(N_{HI})$. However,
such absorbers are probably formed within rapidly expanded
regions and for them both background properties and expansion
rate vary randomly from absorber to absorber.

For the subpopulation of moderately expanded absorbers, we can
combine Eqs. (\ref{delr}), (\ref{br}), (\ref{bb}) and
(\ref{adiab}) and, in principle,  connect the background
temperature with the Doppler parameter of absorbers, $b$, and
their hydrogen column density, $N_{HI}$.
However, reliability and significance of such estimates are
in question. Some restrictions on the intensity of UV background
were discussed in Sec. 4.1\,.

\section{Summary and Discussion.}

In this paper we continue the analysis initiated in Paper I and
Paper II that is based on the statistical description of
Zel'dovich pancakes (DD99, DD04). This approach allows one to
connect the observed characteristics of absorbers with
fundamental properties of the initial perturbations without any
smoothing or filtering procedures, to reveal and to illustrate
the main tendencies of structure evolution. It demonstrates also
the generic origin of absorbers and the Large Scale Structure
observed in the spatial distribution of galaxies at small
redshifts.

We investigate the more representative sample of
$\sim$~6\,000 absorbers what allows us to improve the physical
model of absorbers introduced in Paper I and Paper II and to
obtain reasonable description of physical characteristics of
absorbers. The progress achieved demonstrates again the key role
of the representativity of the observed samples for the
construction of the physical model of absorbers and reveals a
close connection between conclusions and the observational
database. Further progress can be achieved with richer and more
refined sample of observed absorbers.

\subsection{Main results}

Main results of our analysis can be summarized as follows:
\begin{enumerate}
\item
    For suitable parameters of the model (Sec. 4.1), the
    basic observed properties of absorbers and their
    evolution are quite successfully described by the
    statistical model of DM confined structure elements
    (Zel'dovich pancakes) with various evolutionary histories.
    Comparison of independent estimates of the DM characteristics
    of pancakes confirms the self consistency of the physical
    model. This model is in a good agreement with measured
    properties of metal systems (see, e.g. Carswell, Schaye \&
    Kim 2002; Telferet al. 2002; Bergeron \& Herbert-Fort 2005).
\item The PDFs of the DM column density and the distances between
     neighboring absorbers are found to be consistent with the
     Gaussian initial perturbations with the  CDM--like initial
     power spectrum.
\item 
    For the observed range of redshifts the evolution
    of absorbers is close to self--similar one.
    This implies that it leads to slow variations
    of mean absorber characteristics with redshift and retains
    their PDFs.
\item We estimate the shape of the correlation function of
    the initial velocity field what in turn allows us to
    estimate the shape of the initial power spectrum.
    At scales $\geq 0.15h^{-1}$ Mpc both derived
    correlation functions, (\ref{sep_fit}) and (\ref{cor_fit}),
    reproduce the CDM--like one. This means that at such scales
    the power spectrum of initial perturbations is close to the
    standard one (\ref{f1})\,. At smaller scales we see some
    differences between the derived and CDM--like correlation
    functions which depend upon the sample used in the analysis.
\item Analysis of variations of the Doppler parameter, $b$,
    along the line of sight demonstrates existence of
    rapidly expanded regions which can be considered
    as examples of strong negative density perturbations
    of galactic mass scale.
\item Our analysis shows that in the observed range
    of redshifts we can expect slow variations of the
    intensity of UV background radiation and of the
    ionization rate, $\Gamma_{12}$. Our results  are close to
    the estimates of the UV background in Haardt\,\&\,Madau
    (1996), Scott et al. (2002), and Demia\'nski\,\&\,
    Doroshkevich (2004b).
\end{enumerate}

\subsection{Test of the model of absorbers}

The physical model of absorbers introduced in Sec. 3 links the
measured $z$, $b$ and $N_{HI}$ with other physical
characteristics of both gaseous and DM components forming the
observed absorbers. It is important that this 1D model provides
us with the self consistent statistical description of the
Ly--$\alpha$ forest although some parameters of pancakes remain
unknown. Action of these parameters as well as uncertainties in
the available estimates of the background temperature and the UV
background  radiation lead to moderate random scatter of the
derived characteristics of absorbers. Fortunately, actions of
these factors partly compensate each other, what allows us to
obtain reasonable statistical description for majority of
absorbers. The self consistency of this approach is confirmed by
similarity of the functions $\xi_v(q_{rd})$ (\ref{sep_fit}) and
$\xi_v(q)$ (\ref{cor_fit}) and by estimates of the matter
fraction, $f_{abs}$, accumulated by absorbers (\ref{fabs}).
These functions are related to independent characteristics of
absorbers obtained from measurements of their separation and
their DM column density.

For richer absorbers, both the hydrostatic equilibrium of 
compressed matter along the shorter axis of pancakes
and the close link between the gas temperature
and the Doppler parameter, $b$, are confirmed by comparison of
characteristics of the HI and metal systems (see, e.g. Carswell,
Schaye \& Kim 2002; Telferet al. 2002; Simcoe, Sargent \& Rauch
2002, 2004; Boksenberg, Sargent and Rauch 2003; Manning 2002,
2003 a,b; Bergeron \& Herbert-Fort 2005). In particlar, for 191
high resolution metal systems presented in Boksenberg, Sargent
and Rough (2003) differences between the gas temperatures
measured by the Doppler parameters of HI and CIV do not exceed
$\sim 25\%$, what is comparable to the precision of
measurements. Comparison of the Doppler parameters measured for
HI, CIV and OVI (Carswell, Schaye \& Kim 2002) verifies also
their similarity and shows that as a rule the macroscopic
(turbulent) velocities are subsonic. These observational results
strongly support the domination of long--lived  gravitationally
bound and partly relaxed absorbers composed of both DM and
baryonic components.

Numerical simulations show that the line width depends upon the
thermal broadening, the differential Hubble flow and peculiar
velocities and the relative influence of these factors varies from
absorber to absorber (see, e.g., Theuns, Schaye \& Haehnelt 2000;
Schaye 2001). The Hubble flow is more essential for weaker absorbers
and can artificially increase their Doppler parameter. For majority
of absorbers in the considered 1D model the possible contribution of
Hubble flow is naturally linked with the compression or expansion of
pancakes in the transverse directions and depends upon the (unknown)
relative orientation of absorber and the line of sight.  The
available observational data do not allow to discriminate between
the thermal and macroscopic broadening of the lines what increases
the random scatter of our results. To perform such discrimination a
more detailed description of the observed line profiles is required.

One of the important problem facing the high resolution
numerical simulations is the development of the methods for the
more detailed reconstruction of the physical properties and
revealing of links between the DM and baryonic components of
observed absorbers. In particular, this includes the
discrimination of the thermal and macroscopic broadening of
lines, explanation of the surprisingly weak redshift dependence
of the mean Doppler parameter, detection of the complex internal
structure of absorbers as indicated by the observations of metal
systems and so on.

However, now technical limitations restrict facilities of
simulations. Thus, the small box size used eliminates the large
scale part of the power spectrum and decreases the representativity
of simulated sample of absorbers. As was discussed in Paper II and
in Manning (2003 a,b), these factors eliminate the interaction of
large and small scale perturbations and distort characteristics of
the simulated absorbers. Simulations reproduce the observed
transmitted flux and its main features and now they are used mainly
for the surprisingly stable reconstruction of the initial power
spectrum from the flux characteristics (see, e.g., Seljak et al.
2004; McDonald et al. 2004,Viel et al 2004a, b). However, the
analysis of Meiksin, Bryan and Machacek (2001) shows that
simulations have problems with  reproduction of the observed PDFs
for the column density of neutral hydrogen, $N_HI$, and the Doppler
parameter, $b$, and their self similar redshift evolution.

More detailed criticism of the ``Fluctuating Gunn-Petersen
approximation'' and the simulations of the forest can be found
in Manning (2002, 2003 a,b), Paper II and references cited in
these papers.

\subsection{Properties of absorbers}

Analysis of the mean absorbers characteristics performed in Sec.
4 shows that the sample of observed absorbers is composed of
pancakes with various evolutionary histories. We discuss five
main factors that determine evolution of absorbers after their
formation. They are: the transverse expansion and compression of
pancakes, the disruption of structure elements into a system of
high density clouds, the merging of absorbers and the radiative
heating and cooling of compressed gas. The first two factors
change the overdensity of DM and gas but do not change the gas
entropy. Next two factors change both the gas entropy and
overdensity but do not change the DM characteristics.

The sample of observed forest can be naturally divided into
subsamples of adiabatically and shock compressed absorbers
formed by merging. Moreover, about half of adiabatically
compressed absorbers are formed within rapidly expanded regions
where the background temperature is less than mean one. So, the
temperature of absorbers formed within such regions can be also
less than the mean temperature of the background (\ref{bg}).
These results illustrate the influence of some of the factors
mentioned above. However, the slow variation of the mean
characteristics of absorbers and their PDFs with redshift
confirms that we observe the self--similar period of absorbers
evolution when the action of these factors is balanced and
regular variations of the UV background does not distort this
balance.

For shock compressed absorbers, introduction of the DM column
density, $q$ and $\zeta$, and entropy, $S_m\approx S_b$, allows
to discriminate between the systematic and random variations of
their properties. The former ones are naturally related to the
progressive growth with time of the DM column density of
absorbers, $q(z)\,\&\,\zeta(z)$, and they can be described
theoretically. On the other hand the action of random factors
cannot be satisfactorily described by any theoretical model.
However, in the framework of our approach, the joint action of
all random factors is summarized by one random function, $S_b$,
directly expressed through the observed parameters (\ref{sbar}).
These results alleviate the problem of description of absorbers
and, perhaps, the modelling of the Ly-$\alpha$ forest based on
the simulated DM distribution (Viel et al. 2002)\,.

For adiabatically compressed absorbers, the spatial distributions,
entropy and overdensity of baryonic and DM components are
different. Unfortunately at present these characteristics of DM
component cannot be determined from observations with a
reasonable accuracy. For such absorbers the baryonic entropy is
identical to the background one given by (\ref{sbg}) while the
PDFs and the random scatter of observed characteristics are
determined mainly by random variations of the expansion rate and
the background density and temperature. For this subpopulation,
the process of formation and evolution of absorbers should be
investigated more thoroughly.

\subsection{Characteristics of the initial power spectrum}

The initial power spectrum of density perturbations is created
at the period of inflation and its observed determination is
very important for investigations of the early Universe. The
amplitude and the shape of large scale initial power spectrum
are approximately established by investigations of relic
radiation (see. e.g, Spergel et al. 2003, 2006) and the structure 
of the Universe at $z<$~1 detected in large redshift surveys such
as the SDSS (Dodelson et al. 2002; Tegmark et al. 2004) and 2dF
(Percival et al. 2001). The shape of the initial power spectrum
at small scale can be tested at high redshifts where it is not
so strongly distorted by nonlinear evolution (see, e.g., Croft
et al. 2002; Tegmark et al. 2002; McDonald et al. 2004; Seljak
et al. 2004; Zaroubi et al. 2005).

Here we retrieve  the correlation function of initial velocity
field, $\xi_v$, from  direct measurements of the PDFs of
fundamental characteristics of absorbers such as their
separation, $d_{sep}$, and the DM column density, $q$\,. Both
estimates are derived in the same way and result in the same
shape of the correlation function at larger scales, $q\geq
5\cdot 10^{-3}, l_vq\geq 0.15 h^{-1}$ Mpc\,. At these scales the
measured correlation functions coincide with the CDM--like one
(\ref{xiv}) what confirms conclusions of Croft et al. (2002),
Viel et al. (2004b); McDonald et al. (2004) and Zaroubi et al.
(2005) obtained at scales $\geq 1h^{-1}$ Mpc.

At smaller scales the results obtained with analysis of the
absorbers separation, $d_{sep}$, and the DM column density, $q$,
are different and demonstrate some excess of power at scales 150
kpc $\geq l_vq\geq$ 3 kpc. Parameters of these functions and the
excess depend upon the sample of absorbers and for the extended
samples the derived correlation functions become quite similar
to the CDM--like one. However, reliability of this result is in
question due to a probable incompleteness of the extended
samples.

The interpretation of these distortions is not unique because
of very limited available information. As was shown in Sec.
4.2, they are sensitive to the deficit of weaker absorbers
and small separations of absorbers in the sample under
consideration. Therefore  these distortions can be enhanced by
the probable incompleteness of the observed sample created by
the finite resolution of observations,  blending of lines and
approximate character of our analysis. If this explanation is
correct than these factors restrict the presently available
range of investigations to $ l_vq\geq$ 100 kpc\,. Further
progress can be achieved with more refined observations of
absorption spectra of QSOs and with more refined identification
of absorbers in the observed spectra.

In turn, these divergences can be related to special features in
the initial power spectrum at small scales. Recent WMAP
measurements indicate that adiabatic Gaussian perturbations
dominate on large scale (Peiris et al. 2003; Komatsu et al.
2003). However, these results do not preclude deviations from
the standard CDM--like power spectrum at small
scales. In particular, such deviations appear in models of the
one field inflation with a complicated inflation potential (see,
e.g., Ivanov, Naselsky \& Novikov 1994) or multiple fields
inflation (see, e.g., Polarski \& Starobinsky 1995; Turok 1996).
Both models generate adiabatic or isocurvature deviations from
the simple CDM--like power spectrum. More detailed discussion of
such models can be found, for example, in Peiris et al. (2003).

\subsection{Absorbers as elements of the Large Scale
Structure of the Universe}

At redshifts $z\geq 1.7$ the Large Scale Structure is observed
mainly as systems of absorbers in spectra of distant QSOs.
Numerical simulations show that even at such redshifts we can
see also high density filaments and clumps formed by
``galaxies'' and some of them are actually observed in spectra
of QSOs as metal systems, Lyman damped and Lyman limit systems.
However, available observational data do not yet allow one to
characterize statistically properties of such structure elements
(see, e.g., Boksenberg, Sargent\,\&\,Rauch 2003).

At small redshifts, the Large Scale Structure is observed as a
spatial distribution of both galaxies and neutral hydrogen. The
investigation of galaxy distribution in the SDSS DR1
(Doroshkevich, Tucker, Allam \& Way 2004a) results in estimates
of typical parameters of galaxy walls as
\be
\langle q\rangle\approx 0.4,\quad \langle b\rangle\approx 320km/s,
\quad \langle d_{sep}\rangle\approx 60h^{-1} {\rm Mpc}\,.
\label{walls}
\ee
With these data the expected column density of neutral hydrogen
within the typical wall (\ref{NH1}) is $N_{HI}\approx 10^{11}
cm^{-2}$ and even so spectacular object as the 'Greet Wall' does
not manifest itself through absorbers.

Our results indicate the generic link of absorbers and DM
Zel'dovich pancakes and demonstrate that the embryos of walls
could also be seen already at $z\sim 3$. Indeed, for basic
parameters of subpopulation of 1\,370 shock compressed absorbers
with $b\geq 30$ km/s we have
\be
\langle d_{sep}\rangle\approx (50\pm 11)(1+z)^{-2}h^{-1}{\rm
Mpc}\,,
\label{zwal}
\ee
\[
\langle q\rangle\approx (0.4\pm 0.07)(1+z)^{-2}\,,
\]
what is quite similar at z=0 to that given in (\ref{walls}).
This fact indicates that, in principle, such absorbers can be
considered as embryos of wall--like elements of the Large Scale
Structure of the Universe. Of course, such identification of
walls observed in the galaxy distribution with elements of
Ly--$\alpha$ forest is quite arbitrary and ignores the actual
complex evolution of the LSS elements. However, it confirms
generic character of the LSS evolution from richer absorbers
to galaxy walls. The problem deserves further investigation
first of all with more representative numerical simulations.

For the first time poor absorbers at small redshifts were
observed by Morris et al. (1991, 1993) and $\sim 1000$ of such
absorbers were found by Bahcall et al. (1993, 1996) and Jannusi
et al. (1998). Some of these absorbers are identified with halos
of galaxies (see, e.g., Lanzetta et al. 1995; Le Brune, Bergeron
\& Boisse 1996) or galaxy filaments (Penton, Shull \& Stock
2002) but others are situated far from any galaxies. These
observations demonstrate that the space between the LSS elements
-- so called 'voids' -- is not empty and contains essential
fraction of baryonic and DM components of the matter.

More detailed characteristics of absorbers at $z\ll$ 1 are given
in  Penton, Shull \& Stock (2000, 2002); McLin et al. (2002)
where the main absorber properties are found to be similar to
those observed at high redshifts. For 79 absorbers with $12\leq
\lg N_{HI}\leq 15$, $11km/s\leq b\leq 80km/s$ listed in these
papers the mean absorber separation is \be \langle
d_{sep}\rangle\sim (10\pm 3) h^{-1}{\rm Mpc}\,. \label{pss} \ee
Despite the strong difference of many conditions at $z\ll 1$ and
$z\geq 1.5$, these observed characteristics of absorbers are
quite similar to expected ones (\ref{mnsbhs}) extrapolated
to $z=0$.

\subsection{Observed and expected evolution of the
Large Scale Structure}

Comparison of the expected and derived from observations
characteristics of absorbers demonstrates that at the observed
range of redshifts, $1.7\leq z\leq 4.5$ we see the self --
similar period of structure evolution. During this period the
main factors determining the evolution of absorbers such as the
pancake expansion, creation and merging, are balanced what leads
to relatively slow evolution of the mean properties of
absorbers, such as $d_{sep}$ and the DM column density, $q$\,.
This slow evolution is supported by slow regular variations of
the UV background radiation and the ionization rate,
$\Gamma_{12}$\,.

However, at small redshifts, $z\leq 0.5$ the growth of
perturbations and merging of absorbers becomes decelerated due
to the influence of the $\Lambda$--term while expansion and
disruption of absorbers remains important. This means that at
such redshifts the quiet evolution of absorbers is distorted and
we can expect a progressive decrease of linear density of
observed absorbers with the hydrogen column density $N_{HI}\geq
10^{12}cm^{-2}$\,. The variations of the population of observed
absorbers are also modulated by the poorly known variations of
the UV background.

At larger redshifts evolution of DM pancakes is mainly driven
by the shape of the initial power spectrum. For the standard
CDM--like correlation function (\ref{xiv}) we get that
the self--similar evolution takes place at redshifts
\be
 z\leq z_{thr}\sim\sqrt{0.375/q_0}\approx 19\sqrt{10^{-3}
/q_0}\,, \label{evol3} \ee and at $z\geq z_{thr}$ pancakes with
$q\sim q_0$ are more abundant.

However, the observational
test of these expectations is quite problematic because the
observed characteristics of absorbers depend also upon evolution
of the background temperature and UV radiation.

\subsection{Reheating of the Universe}

Recent observations of high redshift quasars with $z\geq$ 5
(Djorgovski et al. 2001; Becker et al. 2001; Pentericci et
al. 2002; Fan et al. 2002, 2003, 2004) provide clear evidence
in favor of the reionization of the Universe at redshifts
$z\sim$~6 when the volume averaged fraction of neutral
hydrogen is found to be $f_H\geq 10^{-3}$ and the
photo ionization rate $\Gamma_\gamma\sim (0.02 - 0.08)\cdot
10^{-12}s^{-1}$ . These results are consistent with
those expected at the end of the reionization epoch which
probably takes place at $z\sim$~6.

These results can be compared with expectations of the
Zel'dovich approximation (DD04). The potential of this approach
is limited since it cannot describe the nonlinear stages of
structure formation and, so, it cannot substitute the high
resolution numerical simulations. However, it describes quite
well many observed and simulated statistical characteristics of
the structure such as the redshift distribution of absorbers and
evolution of their DM column density. This approach does not
depend on the box size, number of points and other limitations
of numerical simulations (see discussion in Paper II) and it
successfully augments them.

This approach shows (DD04) that at $z\sim$~6 only $\sim$ ~3.5\%
of the matter is condensed within the high density clouds which
can be associated with luminous objects. This value can increase
up to $\sim$~5 -- 6\% with more accurate description of the
clouds collapse. The same approach also allows one to estimate
the mass function of structure elements (DD04) at different
redshifts. At $z\sim$ 6, the mean DM mass of the clouds is
expected to be $\langle M_{cl}\rangle\sim 10^{10}M_\odot$ and
majority of clouds have masses between $10^{-3}\langle
M_{cl}\rangle$ and 10 $\langle M_{cl}\rangle$. The formation of
low mass clouds with $M_{cl}\leq 10^6 M_\odot$ is suppressed due
to strong correlation of the initial density and velocity fields
at scales $\leq l_\rho\sim 0.03h^{-1}(q_0/10^{-3})$ Mpc
(\ref{lv}). However, the numerous low mass satellites of large
central galaxies can be formed in the course of disruption of
massive collapsed clouds at the stage of their compression into
thin pancake--like objects (Doroshkevich 1980; Vishniac 1983).
The minimal mass of such satellites was estimated in Barkana,
Haiman \& Ostriker (2001).

This means that the investigation of absorbers observed at high
redshifts should be supplemented by the study of properties of
dwarf {\it isolated} galaxies and discrimination between such
galaxies and dwarf satellites of more massive galaxies. Such
observations seem to be a perspective way to discriminate
between models with one and several types of DM particles.

\subsection*{Acknowledgments}
This work would not have been possible without the important
contribution of M. Rauch and W.L.W. Sargent who provided us with
unpublished spectra of five quasars. We are deeply grateful for
their permission to use their data.  This paper was supported in
part by the Polish State Committee for Scientific Research grant Nr.
1-P03D-014-26 and Russian Found of Fundamental Investigations grant
Nr. 05-02-16302.


\end{document}